\documentclass[showpacs,superscriptaddress,floatfix,aps,prx,notitlepage,reprint]{revtex4-2}
\usepackage{graphicx}
\usepackage{amsfonts,amssymb,amsmath,mathtools}
\usepackage{hyperref,hypcap}
\usepackage{braket}
\usepackage{bm}
\usepackage{color}
\usepackage{bbold}
\usepackage[utf8]{inputenc}
\newcommand{\ii}{\text{i}}
\newcommand{\Tr}{{\rm Tr}}
\newcommand{\mE}{\mathbb{E}}
\newcommand{\mU}{\mathcal{U}}

\newcommand{\mM}{\mathcal{M}}
\newcommand{\mC}{\mathcal{C}}
\newcommand{\mK}{\mathcal{K}}

\newcommand{\br}{\bm r}
\newcommand{\tU}{\tilde{U}}

\newcommand{\Z}{$\mathbb{Z}$}
\newcommand{\Zt}{$\mathbb{Z}_{2}$}

\newcommand{\Ca}{${\cal C}_{0}$ & $0$ & \Z & $0$ & \Z & $0$ & \Z & $0$ & \Z}
\newcommand{\Cb}{${\cal C}_{1}$ & \Z & $0$ & \Z & $0$ & \Z & $0$ & \Z & $0$}

\newcommand{\Ra}{${\cal R}_{0}$ & $0$ & $0$ & $0$ & $2$\Z & $0$ & \Zt & \Zt & \Z}
\newcommand{\Rb}{${\cal R}_{1}$ & \Z & $0$ & $0$ & $0$ & $2$\Z & $0$ & \Zt & \Zt}
\newcommand{\Rc}{${\cal R}_{2}$ & \Zt & \Z & $0$ & $0$ & $0$ & $2$\Z & $0$ & \Zt}
\newcommand{\Rd}{${\cal R}_{3}$ & \Zt & \Zt & \Z & $0$ & $0$ & $0$ & $2$\Z & $0$}
\newcommand{\Ree}{${\cal R}_{4}$ & $0$ & \Zt & \Zt & \Z & $0$ & $0$ & $0$ & $2$\Z}
\newcommand{\Rf}{${\cal R}_{5}$ & $2$\Z & $0$ & \Zt & \Zt & \Z & $0$ & $0$ & $0$}
\newcommand{\Rg}{${\cal R}_{6}$ & $0$ & $2$\Z & $0$ & \Zt & \Zt & \Z & $0$ & $0$}
\newcommand{\Rh}{${\cal R}_{7}$ & $0$ & $0$ & $2$\Z & $0$ & \Zt & \Zt & \Z & $0$}

\newcommand{\kk}[1]{{\textcolor{black}{#1}}}

\DeclareMathAlphabet{\mathitb}{OT1}{cmr}{bx}{sl}

\usepackage{etoolbox}
\makeatletter
\patchcmd{\frontmatter@abstract@produce}
  {\vskip200\p@\@plus1fil
   \penalty-200\relax
   \vskip-200\p@\@plus-1fil}
  {}
  {}
  {}
\makeatother

\begin{document}

\title{Symmetry and Topology of Monitored Quantum Dynamics}

\author{Zhenyu Xiao}
\email{zyxiao@princeton.edu}
\affiliation{International Center for Quantum Materials, Peking University, Beijing 100871, China}
\affiliation{Princeton Quantum Initiative, Princeton University, Princeton, New Jersey 08544, USA}

\author{Kohei Kawabata}
\email{kawabata@issp.u-tokyo.ac.jp}
\affiliation{Institute for Solid State Physics, University of Tokyo, Kashiwa, Chiba 277-8581, Japan}

\date{\today}

\begin{abstract}
The interplay between unitary dynamics and quantum measurements induces diverse phenomena in open quantum systems with no counterparts in closed quantum systems at equilibrium. Here, we generally classify Kraus operators and their effective non-Hermitian dynamical generators, thereby establishing the tenfold classification for symmetry and topology of monitored free fermions. Our classification elucidates the role of topology in measurement-induced phase transitions and identifies potential topological terms in the corresponding nonlinear sigma models. Furthermore, we establish the bulk-boundary correspondence in monitored quantum dynamics: nontrivial topology in spacetime manifests itself as topologically nontrivial steady states and gapless boundary states in Lyapunov spectra, such as Lyapunov zero modes and chiral edge modes, leading to the topologically protected slowdown of dynamical purification.
\end{abstract}

\maketitle

\section{Introduction}
The interplay of unitary dynamics and quantum measurements gives rise to distinctive phenomena in open quantum systems with no analogs in closed quantum systems at equilibrium~\cite{fisher2023a}.
Open quantum dynamics is not described by Hermitian Hamiltonians but by nonunitary Kraus operators~\cite{Nielsen-textbook, Breuer-textbook, Rivas-textbook}, where measurements select a quantum trajectory~\cite{Carmichael-textbook, Plenio-review, Daley-review}.
Unitary dynamics accompanies the propagation of quantum correlations and entanglement, resulting in thermalization~\cite{Eisert-review, Rigol-review}.
By contrast, nonunitary quantum measurements drive the system into nonequilibrium steady states.
Their competition has been shown to induce dynamical quantum phase transitions~\cite{vasseur2019, chan2019, skinner2019, li2018a, li2019, choi2020, gullans2020d, jian2020}, extensively investigated in both theory~\cite{szyniszewski2019, bao2020, tang2020, gullans2020, zabalo2020, goto2020, lavasani2021a, sang2021, ippoliti2021a, szyniszewski2020, lunt2020, li2021b, fidkowski2021, nahum2021, ippoliti2021b, biella2021, bao2021, lu2021a, ippoliti2022, turkeshi2021, jian2021a, minato2022, block2022, zabalo2022, turkeshi2022a, yamamoto2023, lu2023, bulchandani2024, deluca2023, mochizuki2024, yang2023EFT} and experiments~\cite{noel2022, koh2023, hoke2023}.

Monitored free fermions exhibit rich quantum phenomena and have attracted considerable recent interest~\cite{cao2019, nahum2020a, chen2020, alberton2021, jian2022, tang2021a, buchhold2021, muller2022, coppola2022a, kells2023, fleckenstein2022, kawabata2023a, wang2024, merritt2023a, legal2023a, venn2023, szyniszewski2023, behrends2024, jian2023, swann2023, fava2023, loio2023, poboiko2023, chahine2024, poboiko2024, liu_li_xu2024, fava2024, zhenyuxiao2024}.
Measurement-induced phase transitions were numerically found for monitored complex~\cite{chahine2024, poboiko2024} and Majorana~\cite{nahum2020a, merritt2023a} fermions in two and one spatial dimensions, respectively, despite their possible absence for complex fermions in one dimension~\cite{cao2019, alberton2021}.
Symmetry of nonunitary quantum circuits was also studied through a tensor-network framework~\cite{jian2021a}.
While this analysis corresponds to forced measurements with postselected quantum trajectories, subsequent works developed effective nonlinear sigma models for monitored free fermions~\cite{jian2023, fava2023, poboiko2023}, reminiscent of those for the Anderson transitions in disordered electron systems~\cite{anderson1958, abrahams1979, lee1985, beenakker1997, evers2008a}.
From their perturbative analysis, a unique scaling law of entanglement entropy was derived, consistent with numerical calculations~\cite{fava2023}.

Still, the intricate connection between these effective field theory and microscopic nonunitary quantum dynamics has not been fully understood.
Specifically, topological terms can generally be incorporated into nonlinear sigma models, which profoundly influence the Anderson transitions, as exemplified by the quantum Hall transitions~\cite{QHE-textbook, Huckestein-review, Kramer-review}. 
They also underlie the celebrated tenfold classification of topological insulators and superconductors~\cite{schnyder08, kitaev09, ryu10, hasan10, qi11, chiu2016b}.
However, the role of topology in monitored quantum dynamics has remained largely elusive.

In this work, we establish the classification of symmetry and topology for monitored free fermions.
We identify the tenfold symmetry classes for single-particle Kraus operators and associated non-Hermitian dynamical generators (Table~\ref{tab: symmetry}).
Building upon this symmetry classification, we comprehensively classify topology of single-particle monitored quantum dynamics (Table~\ref{tab: topology}).
This classification reveals the role of topology in measurement-induced phase transitions and describes topological terms in the underlying nonlinear sigma models.
Moreover, we demonstrate that nontrivial non-Hermitian topology in spacetime gives rise to topologically nontrivial steady states and anomalous gapless boundary states in Lyapunov spectra,
thereby constituting the bulk-boundary correspondence in monitored quantum dynamics.

The remainder of this work is organized as follows.
In Sec.~\ref{sec: monitored dynamics}, we provide an overview of monitored free-fermion dynamics in continuous time and identify two operators that completely encode the nonunitary time evolution.
In Sec.~\ref{sec: symmetry}, we classify symmetry of the dynamics in terms of Kraus operators, emphasizing that only symmetry preserved under time-ordered products is relevant for the full time evolution.
In Sec.~\ref{sec: topology}, we specify the appropriate gap structures and develop the topological classification based on the two operators, defined in $(d+1)$-dimensional spacetime and in $d$-dimensional space, and demonstrate the equivalence of these formulations.
We further derive the associated bulk-boundary correspondence.
Sections~\ref{sec: num 1D} and~\ref{sec: num 2D} present numerical simulations of monitored Majorana fermions in $1+1$ dimensions and monitored complex fermions in $2+1$ dimensions, respectively.
Finally, we conclude with discussion in Sec.~\ref{sec: discussion}.

\section{Monitored Free fermions}
\label{sec: monitored dynamics}

We study general nonunitary quantum dynamics of monitored free fermions in $d$ spatial dimensions, consisting of both unitary time evolution and continuous measurement.
We divide time into infinitesimal intervals $dt$. 
At times $(\cdots, t-dt, t, t+dt, \cdots)$, we measure the particle number $n_i \coloneqq c_i^{\dagger} c_i$ at each site $i$ ($i = 1,2, \cdots, N$) with fermion creation and annihilation operators $c_i^{\dagger}$'s and $c_i$'s. The measurement at time $t$ is described by the Kraus operator,
\begin{align}
\label{eq:mM}
\mM_t
&\coloneqq
\exp\left\{
\sum_i \left[
(n_i-\langle n_i\rangle_t)\sqrt{\gamma_i}\, dW_t^i \right. \right. \nonumber \\
&\qquad\qquad\qquad\qquad \left. \left. -\gamma_i (n_i-\langle n_i\rangle_t)^2 dt
\right]
\right\}
\nonumber\\
&\propto
\exp\left\{\sum_i \epsilon_i(t) n_i dt\right\},
\end{align}
with
\begin{equation}
\label{eq:epsilon}
\epsilon_i(t) dt
\coloneqq
(2\langle n_i\rangle_t-1)\gamma_i\, dt + \sqrt{\gamma_i}\, dW_t^i .
\end{equation}
Here, $\gamma_i$ is the measurement strength at site $i$, 
$dW_t^i$ is a standard Wiener increment satisfying $\langle dW_t^i\rangle_E=0$ and $\langle dW_t^i dW_t^j\rangle_E=\delta_{ij}dt$, where $\langle\cdot\rangle_E$ denotes the ensemble average, 
and $\langle n_i\rangle_t$ is the expectation value for the state at time $t$.
The stochastic variables $\epsilon_i$'s in Eq.~\eqref{eq:epsilon} encode all possible measurement outcomes.
If we postselect on the measurement record, the resulting distribution of $\epsilon_i$ can be modified into more general forms, e.g., a random distribution independent of $\langle n_i\rangle_t$, known as forced measurement~\cite{jian2020, jian2023, jian2022, chen2020, tang2021a, kawabata2023a, legal2023a, swann2023}.
Our subsequent discussion does not depend on the specific form of the distribution of $\epsilon_i$, and hence applies to scenarios with or without postselection.

The unitary time evolution from $t$ to $t+dt$ is generated by 
\begin{equation}
\mU_t = e^{-\ii h_t dt}, \quad h_t=\sum_{ij} c_i^{\dagger} (h_t)_{ij} c_j
\end{equation}
with a (possibly time-dependent) quadratic Hamiltonian $h_t^{\dagger}=h_t$.
The spatial structure of $h_t$ determines the spatial dimension $d$ of the dynamics.
We also allow $h_t$ to depend on the measurement records, corresponding to feedback control of the quantum state~\cite{wiseman1994,doherty1999}.
The entire time evolution from $0$ to $t$ driven by both measurement and unitary dynamics is characterized by the cumulative Kraus operator $\mathcal{K}_{[0,t]}$, defined as
\begin{equation}
\label{eq:K_0t}
\mathcal{K}_{[0,t]} \coloneqq \mU_{t}\mM_{t}\cdots \mU_{d t}\mM_{d t}
=
T\exp\left\{\int_{0}^{t}\mathcal{H}_s\, ds\right\},
\end{equation}
where $T$ denotes time ordering, and $\mathcal{H}_s$ is 
\begin{equation}
\mathcal{H}_s \coloneqq -\ii \sum_{ij} c_i^{\dagger}(h_s)_{ij}c_j + \sum_i \epsilon_i(s) n_i.
\end{equation}

Since both $\mU_t$ and $\mM_t$ are quadratic \kk{in fermion operators}, the dynamics preserves Gaussianity. 
It is fully encoded by a single-particle propagator $K_{[0,t]}$ associated with $\mathcal{K}_{[0,t]}$ in Eq.~\eqref{eq:K_0t}~\cite{bravyi2004b, zhenyuxiao2024},
\begin{equation}
\label{eq:Kprod}
K_{[0,t]} \coloneqq K_t K_{t-dt}\cdots K_{d t},
\end{equation}
with
\begin{align}
\label{eq:Kt_def}
K_t &\coloneqq e^{H_t dt}, \\
H_t &\coloneqq -\ii h_t + \mathrm{diag}\,\big(\epsilon_1(t),\epsilon_2(t),\ldots,\epsilon_N(t)\big).
\end{align}
Here, $K_t$ acts on a single-particle state $|\psi_t\rangle \coloneqq \sum_i (\psi_t)_i |i\rangle$, with $|i\rangle\coloneqq c_i^{\dagger}|\mathrm{vac}\rangle$.
Over an infinitesimal interval $[t,t+\Delta t]$, the wave function evolves as
\begin{equation}
\label{eq:1p_update}
|\psi_{t+d t}\rangle = K_t|\psi_t\rangle, \qquad
(\psi_{t+d t})_i = \sum_j (K_t)_{ij}(\psi_t)_j .
\end{equation}
Because $K_t$ is generally nonunitary, $\langle\psi_{t+dt}|\psi_{t+dt}\rangle$ does not necessarily equal $\langle\psi_t|\psi_t\rangle$.
In the continuous-time limit, Eq.~\eqref{eq:Kt_def} leads to the stochastic Schr\"odinger equation
\begin{equation}
\label{eq:Lt}
L_t |\psi_t\rangle = 0, \qquad
L_t \coloneqq \partial_t - H_t .
\end{equation}
The relationship between $L_t$ and $K_t$ is analogous to that between Hamiltonians and their transfer matrices in disordered electron systems~\cite{zirnbauer1996, altland1997a, beenakker1997, evers2008a}, where the temporal direction is replaced with the spatial direction. 

Although $L_t$ and $K_{[0,t]}$ are single-particle operators, they completely determine the evolution of a many-body fermionic Gaussian state. 
For instance, suppose that the initial state is a Slater determinant built from $m$ single-particle orbitals, \begin{equation}
|\Psi_0\rangle
=
\prod_{\alpha=1}^{m}
\left[\sum_i (\psi_{\alpha}(0))_i\, c_i^{\dagger}\right]
|\mathrm{vac}\rangle, 
\end{equation}
with $|\mathrm{vac}\rangle$ being the empty state.
The unnormalized state $|\tilde{\Psi}_t\rangle$ at time $t$ remains a Slater determinant, with each orbital propagated by $K_{[0,t]}$, given as 
\begin{equation}
|\tilde{\Psi}_t\rangle
=
\prod_{\alpha=1}^{m}
\left[\sum_i (\psi_{\alpha}(t))_i\, c_i^{\dagger}\right]
|\mathrm{vac}\rangle 
\end{equation}
with $(\psi_{\alpha}(t))_i
=
\sum_j (K_{[0,t]})_{ij}(\psi_{\alpha}(0))_j$. 
The time evolution of the Gaussian mixed 
states can also be determined by $K_{[0,t]}$ similarly~\cite{zhenyuxiao2024}.
Furthermore, many-body characteristics of the monitored dynamics, such as purification and steady-state properties, can therefore be extracted directly from $L_t$ or $K_{[0,t]}$ (see Sec.~\ref{subsec: Topology Lt}).

\section{Symmetry Classification}
\label{sec: symmetry}
The single-particle Kraus operator $K_{[0,t]}$ completely captures the conditional time evolution. 
We therefore classify the monitored dynamics by symmetry of $K_{[0,t]}$, which is expected to control key physical properties of monitored free fermions, including topological phenomena and measurement-induced phase transitions.
Because $K_{[0,t]}$ is a time-ordered product of $K_t$ at different time slices [see Eq.~\eqref{eq:Kprod}], its symmetry group is determined by, and in general no larger than, that of the individual factors $K_t$.
Moreover, $K_t$ inherently incorporates spacetime randomness arising from spatial disorder and the temporal noise intrinsic to quantum measurements, manifested in the stochastic fields $\epsilon_i(t)$ [Eq.~\eqref{eq:epsilon}].
As a result, $K_t$'s at different times generally do not commute. 
Only symmetry preserved under multiplication survives and is promoted to symmetry of $K_{[0,t]}$.

Symmetry of $K_t$ is also determined by symmetry of $H_t$ in Eq.~\eqref{eq:Kt_def}, which is generically neither Hermitian nor anti-Hermitian.
For example, if $H_t$ obeys a transposition symmetry $H_t=\mathcal{U} H_t^{\rm T}\mathcal{U}^{-1}$, then $K_t$ satisfies $\mathcal{U} K_t^{\rm T}\mathcal{U}^{-1}=K_t$.
However, this symmetry is not preserved by the time-ordered product: Eq.~\eqref{eq:Kprod} implies
$\mathcal{U} K_{[0,t]}^{\rm T}\mathcal{U}^{-1}=K_{\Delta t}\cdots K_{t-\Delta t}K_t \neq K_{[0,t]}$
in general.
Similarly, an inversion-type (sublattice) symmetry of $H_t$, $H_t=-\mathcal{U} H_t \mathcal{U}^{-1}$, implies $\mathcal{U}K_t^{-1}\mathcal{U}^{-1}=K_t$, which is also generically lost under multiplication.
In contrast, if each $H_t$ respects time-reversal symmetry $\mathcal{T} H_t^{*}\mathcal{T}^{-1}=H_t$, then $K_t$ obeys $\mathcal{T} K_t^{*}\mathcal{T}^{-1}=K_t$.
Consequently, the product $K_{[0,t]}$ satisfies
\begin{align}
\mathcal{T} K_{[0,t]}^{*}\mathcal{T}^{-1}
&=
\left(\mathcal{T} K_t^{*}\mathcal{T}^{-1}\right)
\left(\mathcal{T} K_{t-\Delta t}^{*}\mathcal{T}^{-1}\right)
\cdots
\left(\mathcal{T} K_{\Delta t}^{*}\mathcal{T}^{-1}\right)
\nonumber\\
&=
K_t K_{t-\Delta t}\cdots K_{\Delta t}
=
K_{[0,t]} ,
\end{align}
thereby preserving the same time-reversal symmetry.

We enumerate candidate internal symmetries of non-Hermitian operators and determine, as above, which of them are preserved under time-ordered products.
In particular, we identify the following spacetime-internal symmetries that survive multiplication:
\begin{align}
    \mathcal{T} K^{*}_t \mathcal{T}^{-1} &= K_t \quad \left( \mathcal{T}\mathcal{T}^{*} = \pm 1 \right), \label{eq: TRS-K} \\
    \mathcal{C}\,( K^{\rm T}_t )^{-1} \mathcal{C}^{-1} &= K_t \quad \left( \mathcal{C}\mathcal{C}^{*} = \pm 1 \right), \label{eq: PHS-K} \\
    \Gamma\,( K^{\dag}_t )^{-1} \Gamma^{-1} &= K_t \quad \left( \Gamma^2 = 1 \right), \label{eq: CS-K}
\end{align}
where $\mathcal{T}$, $\mathcal{C}$, and $\Gamma$ are unitary operators.
Notably, the symmetry in Eq.~\eqref{eq: PHS-K} combines the transposition and inversion-type symmetries discussed earlier.
Either transposition or inversion alone reverses the temporal ordering and therefore does not define a symmetry of a generic quantum trajectory, whereas their combination yields Eq.~\eqref{eq: PHS-K}.

When $K_t$ satisfies Eqs.~\eqref{eq: TRS-K}--\eqref{eq: CS-K}, the corresponding $H_t$ in Eq.~\eqref{eq:Kt_def} and the non-Hermitian dynamical generator $L_t$ in Eq.~\eqref{eq:Lt} obey
\begin{align}
   \mathcal{T} H^{*}_t \mathcal{T}^{-1} &= H_t,  &   \mathcal{T} L^{*}_t \mathcal{T}^{-1} &= L_t \quad \left( \mathcal{T}\mathcal{T}^{*} = \pm 1 \right), \label{eq: TRS-L} \\
    \mathcal{C} H^{\rm T}_t \mathcal{C}^{-1} &= -H_t,  &  \mathcal{C} L^{\rm T}_t \mathcal{C}^{-1} &= -L_t \quad \left( \mathcal{C}\mathcal{C}^{*} = \pm 1 \right), \label{eq: PHS-L} \\
     \Gamma H^{\dag}_t \Gamma^{-1} &= -H_t , &  \Gamma L^{\dag}_t \Gamma^{-1} &= -L_t \quad \left( \Gamma^2 = 1 \right). \label{eq: CS-L}
\end{align}
The correspondence between symmetry of $H_t$ and $K_t$ follows directly from Eq.~\eqref{eq:Kt_def}.
Moreover, $\partial_t$ is real and anti-Hermitian, i.e., $\partial_t^{*}=\partial_t$, $\partial_t^{\dag}=-\partial_t$, and $\partial_t^{\rm T}=-\partial_t$.
Since $\mathcal{T}$, $\mathcal{C}$, and $\Gamma$ are time independent, they commute with $\partial_t$. Therefore, we have
\begin{equation}
    \mathcal{T} \partial_t^{*} \mathcal{T}^{-1} = \partial_t,\quad
    \Gamma \partial_t^{\dag} \Gamma^{-1} = - \partial_t,\quad
    \mathcal{C} \partial_t^{\rm T} \mathcal{C}^{-1} = - \partial_t .
\end{equation}
Consequently, $\partial_t$ is compatible with Eqs.~\eqref{eq: TRS-L}--\eqref{eq: CS-L}, and $L_t \coloneqq \partial_t - H_t$ inherits the same symmetry class as $H_t$.
Within the 38-fold classification of non-Hermitian operators~\cite{Bernard-LeClair-02, kawabata2019}, these symmetries are called time-reversal, particle-hole, and chiral symmetries, respectively, constituting the tenfold classification in Table~\ref{tab: symmetry}.
Notably, although we label the symmetry classes using the same Altland-Zirnbauer notation as for Hermitian systems (e.g., classes A, D, and DIII), in the non-Hermitian setting these labels carry a different meaning and should not be conflated with their Hermitian counterparts.

While our symmetry classification is consistent with the tensor-network framework~\cite{jian2022}, it encompasses more generic nonunitary quantum dynamics, including those arising from Born measurements.
Although Eqs.~(\ref{eq: TRS-K}) and (\ref{eq: TRS-L}) are referred to as time-reversal symmetry for notational convenience, they do not correspond to the physical time-reversal operation.
Indeed, physical time-reversal symmetry, $\mathcal{T} K^{*}_t \mathcal{T}^{-1} = K_{-t}$, no longer serves as internal symmetry in spacetime, and is thus irrelevant to the monitored quantum dynamics.
Moreover, chiral symmetry in Eqs.~(\ref{eq: CS-K}) and (\ref{eq: CS-L}) is equivalent to (pseudo-)unitarity of transfer matrices~\cite{beenakker1997}.

\begin{table*}[t]
	\centering
	\caption{Tenfold symmetry classification of single-particle Kraus operators $K_{[0:t]}$, associated non-Hermitian dynamical generators $L_t$, and $\bar{H}_t$ defined by $K_{[0:t]} \eqqcolon e^{\bar{H}_t t}$ based on time-reversal symmetry (TRS), particle-hole symmetry (PHS), and chiral symmetry (CS).
    Their classifying spaces are shown in the last three columns.
    The column ``$L_t$" also shows the symmetry classes of the corresponding Hermitian Hamiltonians with the same classifying spaces in the brackets. 
    $L_t$ and $\bar{H}_t$ share the same symmetry but form different classifying spaces because of different gap structures.}
	\label{tab: symmetry}
     \begin{tabular}{ccccccc} \hline \hline \noalign{\vskip 0.4ex} 
  ~~Class~~ & ~~TRS $\mathcal{T}$~~ & ~~PHS $\mathcal{C}$~~ & ~~CS $\Gamma$~~ &~~$L_t$~~ & ~~~$\bar{H}_t$~~~  & 
    $K_{[0:t]}$  \\  [0ex] \hline
    A & $0$ & $0$ & $0$ & $\mathrm{U}(N) \cong \mathcal{C}_1$ (AIII) & $\mathcal{C}_0$ &
    $\mathrm{GL} \left( N, \mathbb{C} \right)/\mathrm{U} \left( N \right)$ 
    \\
    AIII & $0$ & $0$ & $1$ & $\mathrm{U} \left( 2N \right)/\mathrm{U} \left( N \right) \times \mathrm{U} \left( N \right) \cong \mathcal{C}_0$ (A) & $\mathcal{C}_1$ & 
   $\mathrm{U} \left( N, N \right)/\mathrm{U} \left( N \right) \times \mathrm{U} \left( N \right)$ 
    \\ \hline
    AI & $+1$ & $0$ & $0$ &  $\mathrm{O}(N) \cong \mathcal{R}_1$ (BDI) & $\mathcal{R}_0$ & 
    $\mathrm{GL} \left( N, \mathbb{R} \right)/\mathrm{O} \left( N \right)$ 
    \\
    BDI & $+1$ & $+1$ & $1$ &  $\mathrm{O}(2N)/\mathrm{U}(N) \cong \mathcal{R}_2$ (D) & $\mathcal{R}_1$   &$\mathrm{O} \left( N, N \right)/\mathrm{O} \left( N \right) \times \mathrm{O} \left( N \right)$ 
    \\
    D & $0$ & $+1$ & $0$ & $\mathrm{U}(2N)/\mathrm{Sp}(N) \cong \mathcal{R}_3$ (DIII) & $\mathcal{R}_2$ &$\mathrm{O} \left( N, \mathbb{C} \right)/\mathrm{O} \left( N \right)$ 
    \\
    DIII & $-1$ & $+1$ & $1$ & $\mathrm{Sp} \left( 2N \right)/\mathrm{Sp} \left( N \right) \times \mathrm{Sp} \left( N \right) \cong \mathcal{R}_4$ (AII)   &  $\mathcal{R}_3$ & 
    $\mathrm{O}^{*} \left( 2N \right)/\mathrm{U} \left( N \right)$ 
    \\
    AII & $-1$ & $0$ & $0$ &  $ \mathrm{Sp} \left( N \right) \cong \mathcal{R}_5$ (CII) & $\mathcal{R}_4$& $\mathrm{U}^{*} \left( 2N \right)/\mathrm{Sp} \left( N \right)$ 
    \\
    CII & $-1$ & $-1$ & $1$ &  $\mathrm{Sp} \left( N \right) / \mathrm{U} \left( N \right) \cong \mathcal{R}_6$ (C) & $\mathcal{R}_5$ &$\mathrm{Sp} \left( N, N \right)/\mathrm{Sp} \left( N \right) \times \mathrm{Sp} \left( N \right)$ 
    \\
    C & $0$ & $-1$ & $0$ &  $\mathrm{U} \left( N \right) / \mathrm{O} \left( N \right) \cong \mathcal{R}_7$ (CI) & $\mathcal{R}_6$ & $\mathrm{Sp} \left( N, \mathbb{C} \right)/\mathrm{Sp} \left( N \right)$ 
    \\
    CI & $+1$ & $-1$ & $1$ & $\mathrm{O} \left( 2N \right)/\mathrm{O} \left( N \right) \times \mathrm{O}(N) \cong \mathcal{R}_0$ (AI) & $\mathcal{R}_7$ & $\mathrm{Sp} \left( N, \mathbb{R} \right)/\mathrm{U} \left( N \right)$ 
    \\ \hline \hline
  \end{tabular}
\end{table*}

\section{Topology of monitored free fermions}
\label{sec: topology}
\subsection{Topological classification}
\label{subsec: Topology Lt}

We characterize topology of non-Hermitian dynamical generators $L_t$.
Two $L_t$'s are topologically equivalent if they can be deformed into each other while preserving symmetry and a certain gap structure; otherwise, they are topologically distinct. 
Their topology is captured by the homotopy group of their classifying spaces, dependent on the form of the gap. 
In the present setting, intrinsic spacetime randomness generally precludes a conventional spectral gap.
This situation is reminiscent of disordered topological insulators, where quenched disorder obscures the energy gap, yet topology remains stable due to the localization of in-gap states (see, e.g., Refs.~\cite{prodan2010, wangtong21}).

To establish a well-defined topological classification, we require $L_t$ to exhibit a mobility gap at zero.
We will show that this mathematical condition also implies that the dynamics lies in a purifying phase (see a proof below), characterized by a finite purification time $\tau_P=\mathcal{O}(1)$.
The purification time $\tau_P$ is defined by the asymptotic decay of an entropy measure $S(t)$ of mixed states, via 
\begin{equation}
    \frac{1}{\tau_P} \coloneqq -\lim_{t\to\infty} \frac{\ln S(t)}{t}.
\end{equation}
As discussed earlier, $L_t$ and $K_{[0:t]}$ encode the complete information of the conditional dynamics and therefore determine $\tau_P$.
Starting from the maximally mixed state $\rho_0 \propto 1$, the evolved state remains Gaussian and can be written as
\begin{equation}
\rho_t \propto e^{\sum_{ij} 2P_{ij} c_i^{\dag} c_j}, \quad e^{2P} \coloneqq K_{[0:t]}K_{[0:t]}^{\dag}.
\end{equation}
The $\alpha$-R\'enyi entropy $S_{\alpha}$ of a fermionic Gaussian state $\rho_t$ is given by~\cite{cheong2004}
\begin{equation}
S_{\alpha} \coloneqq (1-\alpha)^{-1} \ln \Tr\,\rho_t^{\alpha}
= \sum_{i = 1}^N f_{s\alpha}(\lambda_i),
\end{equation}
where $e^{\lambda_i(t)}$'s are the singular values of $K_{[0:t]}$, and
$f_{s\alpha}(\lambda) \coloneqq
(1-\alpha)^{-1} \ln\!\left[ (1+e^{2\lambda})^{-\alpha} + (1+e^{-2\lambda})^{-\alpha} \right]$.
For $\alpha>1$ and $|\lambda|\gg 1$, we have $f_{s\alpha}(\lambda) \sim \alpha/(\alpha-1)\, e^{-2|\lambda|}$, and the long-time decay of $S_{\alpha}$ is controlled by the smallest growth rate of $|\lambda_i(t)|$.
Therefore, the purification rate can be expressed as
\begin{equation} \label{eq: tau eta}
    \tau_P^{-1} = 2 \min_i|\eta_i|,\quad
    \eta_i \coloneqq \lim_{t \rightarrow \infty} \frac{\langle \lambda_i(t) \rangle_E}{t},
\end{equation}
where $\eta_i$'s are the Lyapunov exponents of $K_{[0:t]}$.


With this in mind, we show that a mobility gap of $L_t$ implies finite $\tau_P$ by proving the contrapositive.
Assume that $\tau_P$ diverges. 
Then, Eq.~\eqref{eq: tau eta} implies that at least one Lyapunov exponent vanishes, i.e., $\eta_{\alpha}=0$ for some $\alpha$.
By the definition of Lyapunov exponents, there exists an initial state $|\phi_0\rangle$ such that the propagated state $|\phi_t\rangle \eqqcolon K_{[0:t]}|\phi_0\rangle$ has the norm growth
$\| |\phi_t\rangle \| = e^{\lambda_{\alpha}(t)} \| |\phi_0\rangle \|$
with $\lim_{t\to\infty}\lambda_{\alpha}(t)/t = 0$.
Equivalently, the sequence $(|\phi_0\rangle,|\phi_{\Delta t}\rangle,\ldots,|\phi_t\rangle)$ solves $L_t|\phi_t\rangle=0$ while remaining extended along the temporal direction, indicating the absence of a mobility gap at zero.
This is also reminiscent of disordered static Hamiltonians, where the localization length of an eigenstate satisfies $\xi = 1/\min_n|\gamma_n|$, with the Lyapunov exponents $\gamma_n$'s of the corresponding transfer matrix~\cite{slevin14}.
The divergence of $\xi$ signals a spatially extended state and is accompanied by $\min_n|\gamma_n|=0$.

\begin{table*}[t]
	\centering
	\caption{Tenfold topological classification of single-particle monitored quantum dynamics $L$ in $d$ spatial dimensions and one temporal dimension.}
	\label{tab: topology}
     \begin{tabular}{cccccccccc} \hline \hline
    ~~Class~~ & ~~~~~~ & ~~$d+1=1$~~ & ~~$d+1=2$~~ & ~~$d+1=3$~~ & ~~$d+1=4$~~ & ~~$d+1=5$~~ & ~~$d+1=6$~~ & ~~$d+1=7$~~ & ~~$d+1=8$~~ \\ \hline
    A & \Cb \\
    AIII & \Ca \\ \hline
    AI & \Rb \\
    BDI & \Rc \\
    D & \Rd \\
    DIII & \Ree \\
    AII & \Rf \\
    CII & \Rg \\
    C & \Rh \\
    CI & \Ra \\ \hline \hline
  \end{tabular}
\end{table*}

Since localized in-gap states do not affect the bulk topology, we assume a spectral gap at zero to proceed with the topological classification.
Owing to the gap, $L_t$ can be continuously deformed into a unitary operator $U$ while preserving all the symmetries of $L_t$ in Eqs.~\eqref{eq: TRS-L}--\eqref{eq: CS-L}.
Concretely, we perform the polar decomposition $L_t=UP$, where $U$ is unitary and $P$ is positive definite, 
\kk{as guaranteed by the spectral-gap condition}
$\det L_t \neq 0$ (see Appendix~\ref{sec: Lt} for details).
The resulting unitary operator $U$ defines a classifying space, denoted by $\mathcal{C}_s$ ($s=0, 1$) or $\mathcal{R}_s$ ($s=0, 1, \cdots, 7$);
see Table~\ref{tab: symmetry}.
For example, in the absence of any symmetry (class A), $U$ can be a generic unitary matrix in $\mathrm{U}(N)$, and the corresponding classifying space is $\mathcal{C}_1$. We identify the classifying spaces of $L_t$ in the remaining symmetry classes in Appendix~\ref{sec: Lt}.

We thus develop the tenfold topological classification of non-Hermitian dynamical generators $L_t$ in $\left( d+1 \right)$-dimensional spacetime in Table~\ref{tab: topology}.
From the homotopy perspective, it is captured by $\pi_{0}\,( \mathcal{C}_{s - (d+1)})$ or $\pi_{0}\,( \mathcal{R}_{s-(d+1)} )$.
It applies to both Born and forced measurements since the underlying classifying spaces are common.
Analogous to equilibrium topological insulators~\cite{schnyder08, kitaev09, ryu10, hasan10, qi11, chiu2016b}, Table~\ref{tab: topology} hosts twofold or eightfold periodicity with respect to spacetime dimensions $d+1$, arising from the Bott periodicity~\cite{Karoubi}.
The tenfold way in Table~\ref{tab: topology} describes possible topological terms in the effective nonlinear sigma models~\cite{jian2023, fava2023, poboiko2023}, elucidating topological measurement-induced phase transitions.
It differs from the classification of Kraus operators $K_t$ or Lindbladians, which reduces to a different tenfold way~\cite{lieu2020, altland2021, sa2023, kawabata2023PRXQ, nakagawa2024}.
In zero spatial dimension $d=0$, topology can protect dynamical quantum criticality with divergent purification time~\cite{zhenyuxiao2024}, akin to disordered chiral-symmetric~\cite{dyson1953, brouwer1998, mondragonShem2014, altland2014a} and nonreciprocal~\cite{Hatano-Nelson-96, Hatano-Nelson-97} wires \kk{(see Appendix~\ref{sec: 0D} for details).}

\subsection{Bulk-boundary correspondence}
\label{subsec: BBC}
We further demonstrate that the spacetime topology of $L_t$ induces nontrivial topology in steady states and anomalous boundary modes in the Lyapunov spectra.
We establish this bulk-boundary correspondence by formulating an equivalent characterization of the topology of the dynamics in terms of the cumulative Kraus operator $K_{[0,t]}$.
We define a non-Hermitian operator $\bar{H}_t$ through 
\begin{equation}
K_{[0,t]} \eqqcolon e^{\bar{H}_t t},
\end{equation}
where $\bar{H}_t$ is an effective time average of $H_t$ [see also Eq.~\eqref{eq:Kt_def}] over the interval $[0,t]$, and inherits all the symmetries of $H_t$ in Eqs.~\eqref{eq: TRS-L}--\eqref{eq: CS-L}.
Let $z_i$'s be complex eigenvalues of $\bar{H}_t$.
As discussed before, we characterize topology within purifying phases, which also imposes gap constraints on the spectrum of $\bar{H}_t$.
According to the Oseledets theorem~\cite{furstenberg1960products}, the Lyapunov exponents are 
\begin{equation}
    \eta_n = \mathrm{Re}\,z_n \quad \left( t \to \infty \right).
\end{equation}
Then, finite $\tau_P = 2 / \min_n |\eta_n|$ requires ${\rm Re}\,z_i \neq 0$,
referred to as a real line gap~\cite{kawabata2019}.

Due to this distinct gap structure (a real line gap), $\bar{H}_t$ can be continuously deformed into a Hermitian Hamiltonian that preserves symmetry.
Concretely, we perform a Schur decomposition 
\begin{equation}
\bar{H}_t = Q r Q^{\dagger}, 
\end{equation}
where $Q$ is unitary and $r$ is upper triangular.
The diagonal entries of $r$ coincide with the eigenvalues of $\bar{H}_t$, i.e., $r_{ii}=z_i$.
We order them such that ${\rm Re}\,r_{ii}>0$ for $1\le i\le p$ and ${\rm Re}\,r_{ii}<0$ for $p+1\le i\le N$, where $p$ and $N-p$ denote the numbers of eigenvalues with positive and negative real parts, respectively.
We then define $\mE \coloneqq {\rm diag}(1_{p},-1_{N-p})$ and consider the continuous path
\begin{equation}
    \bar{H}_t(\lambda) \coloneqq Q\big[\lambda \mE + (1-\lambda) r\big]Q^{\dagger} \quad \left( 0 \leq \lambda \leq 1 \right).
\end{equation}
Along this deformation, the real line gap at ${\rm Re}\,z=0$ remains open, and $\bar{H}_t$ is smoothly deformed into the Hermitian matrix $Q\mE Q^{\dagger}$.
In Appendix~\ref{sec: Ht}, we further show that $Q\mE Q^{\dagger}$ preserves all the symmetries of $\bar{H}_t$ inherited from Eqs.~\eqref{eq: TRS-L}--\eqref{eq: CS-L}.

The classifying space of $\bar{H}_t$ can therefore be read off from the flattened Hermitian representative $Q\mE Q^{\dagger}$.
Notably, $\bar{H}_t$ shares the same internal symmetries as $L_t$ but is associated with a different classifying space because of a different gap structure.
As an illustration, consider the case with no symmetry (class A), in which the classifying space of $L_t$ is $\mathcal{C}_1$.
Then, $Q$ is a generic element of $\mathrm{U}(N)$.
Because $\mE={\rm diag}(1_p,-1_{N-p})$ is invariant under the right multiplication $Q\mapsto QV$ by a block-diagonal matrix $V = {\rm diag}(V_p,V_{N-p})$ with $V_p\in \mathrm{U}(N)$ and $V_{N-p}\in \mathrm{U}(N-p)$, the flattened Hamiltonian $Q\mE Q^{\dagger}$ depends only on the coset space $\mathrm{U}(N)/\mathrm{U}(p)\times \mathrm{U}(N-p)$, denoted by $\mathcal{C}_0$.

More generally, if the classifying space of $L_t$ is $\mathcal{C}_s$ (or $\mathcal{R}_s$), the associated $\bar{H}_t$ is classified by $\mathcal{C}_{s-1}$ (or $\mathcal{R}_{s-1}$) (Table~\ref{tab: symmetry}; see also Appendix~\ref{sec: Ht}).
Unlike $L_t$, $\bar{H}_t$ is defined in $d$ spatial dimensions, and its topology is characterized by $\pi_0(\mathcal{C}_{(s-1)-d})$ or $\pi_0(\mathcal{R}_{(s-1)-d})$.
Thus, $L_t$ and $\bar{H}_t$ fall into the same topological classification, implying that they capture the same topological aspect of the monitored dynamics.
We further show that their topological invariants coincide \kk{(see Appendix~\ref{sec: Green} for details).} 
This equivalence and dimensional reduction are analogous to the scattering-matrix formulation of topological insulators and superconductors, where the topological invariant of a $d$-dimensional Hamiltonian can be encoded in an associated $(d-1)$-dimensional scattering matrix~\cite{fulga2011a}.
The possible nonlocality of $\bar{H}_t$ is irrelevant to the topological classification~\cite{lepori17}.

\begin{figure*}[th]
  \centering
  \includegraphics[width=1\linewidth]{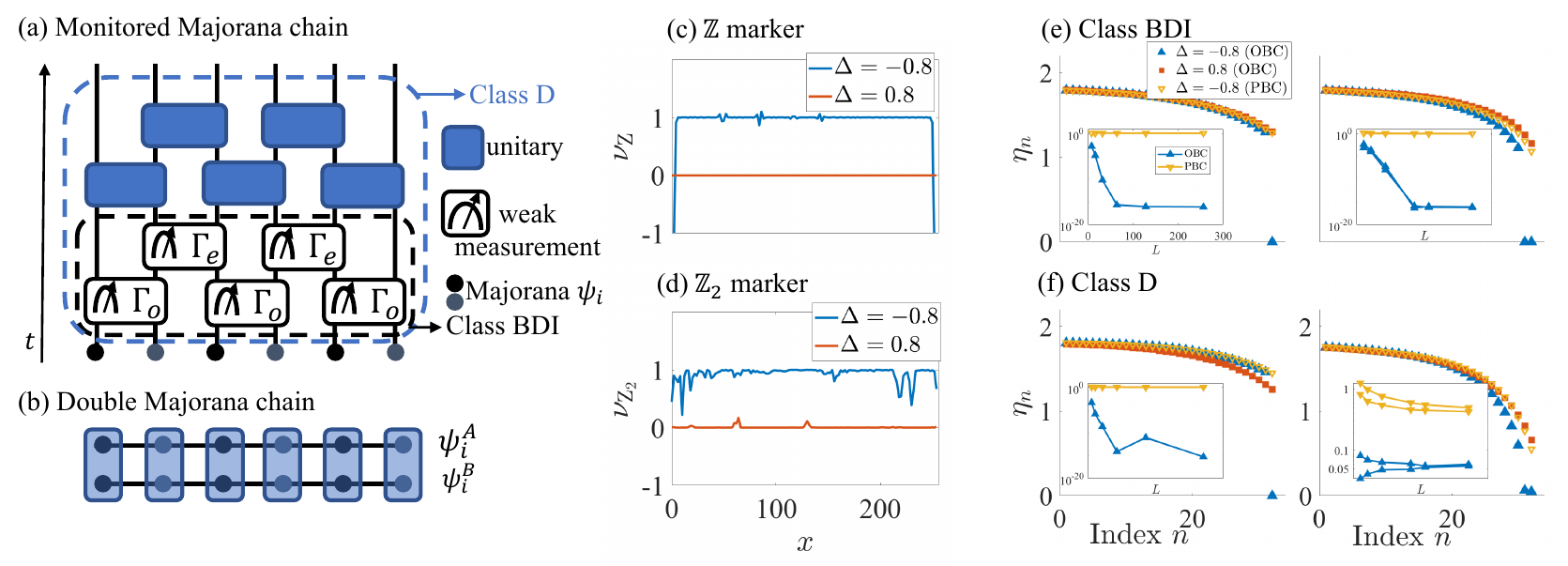}
  \caption{(a)~Monitored dynamics of a Majorana chain generated by repeating the operations inside the black (blue) dashed lines for class BDI (D).
  (b)~Monitored dynamics of the double Majorana chain generated by the operations in subfigure\,(a) on individual chains and the unitary gates coupling the two chains. 
  (c)~$\mathbb{Z}$ and (d)~$\mathbb{Z}_2$ topological markers for the steady states in classes BDI and D, respectively. 
  The fluctuations are due to the spatial randomness in $K_{[0, t]}$.  
  (e), (f)~Lyapunov spectra of the monitored single (left) and double (right) Majorana chains in classes (e)~BDI and (f)~D. 
  Insets:~smallest (or two smallest) positive Lyapunov exponent(s) as a function of the system size. 
  Due to particle-hole symmetry, Lyapunov exponents appear in opposite-sign pairs $(\eta_i, -\eta_i)$ and are shown only for $\eta_i \geq 0$.} 
  \label{fig: chain}
\end{figure*}

In the topological phase, while $\bar{H}_t$ is gapped under the periodic boundary conditions, it becomes gapless under the open boundary conditions due to the emergence of anomalous boundary states, resulting in the topologically protected divergence of $\tau_P$.
Moreover, topology of $\bar{H}_t$ manifests itself in the steady-state correlation function 
\begin{equation}
C_{ji} \coloneqq \braket{\Psi_S | c_i^{\dag} c_j  | \Psi_S} - \frac{\delta_{ij}}{2}.
\end{equation}.
For a generic initial density matrix, the steady state $\Ket{\Psi_S}$ is the many-body eigenstate of $K_{[0:t]}$ with the largest norm of the eigenvalue.
Equivalently, in terms of $\bar{H}_t$, it is obtained by filling the single-particle modes with $\mathrm{Re}\,z_n>0$:
\begin{equation}
\label{eq:steady_slater_R}
|\Psi_S\rangle \propto \prod_{n=1}^{p}\left(\sum_i c_i^{\dag} (\vec{R}_n)_i\right)|\mathrm{vac}\rangle ,
\end{equation}
where $\vec{R}_n$'s are right eigenvectors of $\bar{H}_t$ with eigenvalues $z_n$'s satisfying $\mathrm{Re}\,z_n>0$.

Within a real line gap and fixed symmetry class, $\bar{H}_t$ is continuously deformable to the correlation matrix $C$, implying that they share the same topology.
Diagonalize $\bar{H}_t$ as $\bar{H}_t = S\Lambda S^{-1}$, where the columns of $S$ are the right eigenvectors, $S_{i n} \coloneqq (\vec{R}_n)_i$, and $\Lambda_{nn}\coloneqq z_n$.
Perform a QR decomposition $S = Q' r'$, with $Q'$ unitary and $r'$ upper triangular.
Then, we have
$\bar{H}_t = Q'\,(r'\Lambda r'^{-1})\,Q'^{\dagger}$,
which is precisely a Schur decomposition since $r'\Lambda r'^{-1}$ is upper triangular.
Consequently, $Q'$ coincides with the Schur unitary $Q$ up to a block-diagonal gauge rotation within the positive- and negative-real-part subspaces, i.e., $Q' = Q\,\tilde{U}$ with $\tilde{U}\in \mathrm{U}(p)\times \mathrm{U}(N-p)$.
While the modes $\sum_i c_i^{\dag}(\vec{R}_n)_i$ in Eq.~\eqref{eq:steady_slater_R} are not orthonormal in general, the same Slater determinant can be expressed using an orthonormal single-particle basis obtained from the QR decomposition (see, e.g., Ref.~\cite{cao2019}):
$|\Psi_S\rangle \propto \prod_{n=1}^{p}\left(\sum_i c_i^{\dag} (Q')_{i n}\right)|\mathrm{vac}\rangle$.
For such a Slater determinant, the correlation function takes the projector form
$C = \frac{1}{2}Q'\mE Q'^{\dagger}$, with $\mE \coloneqq {\rm diag}(1_p,-1_{N-p})$.
Using $Q' = Q\tilde{U}$ and the fact that $\tilde{U}$ commutes with $\mE$, we further obtain $C = \frac{1}{2}Q\mE Q^{\dagger}$.
Together with the previous flattening deformation from $\bar{H}_t$ to $Q\mE Q^{\dagger}$, this establishes a continuous deformation from $\bar{H}_t$ to $C$ without closing the real line gap, and then $\bar{H}_t$ and $C$ share the same topology.
This finding also provides a practical route to computing topological invariants of $L_t$ or $\bar{H}_t$, whose direct numerical characterization can be costly.
Instead, we can evaluate the corresponding real-space invariants from $C$ using standard tools developed for disordered topological insulators~\cite{prodan2010, hannukainen2022},
as exemplified in Secs.~\ref{sec: num 1D} and \ref{sec: num 2D}.

\section{Topological monitored dynamics in \texorpdfstring{$1+1$}{1+1} dimensions}
\label{sec: num 1D}

\subsection{Monitored Majorana fermions in class BDI}

As an illustrative example, we consider a measurement-only circuit dynamics of a Majorana chain, generated by iterated weak Born measurements on neighboring Majorana pairs [Fig.~\ref{fig: chain}\,(a)].
Although our general discussion in Sec.~\ref {sec: monitored dynamics} is formulated in continuous time, the classification of symmetry and topology directly applies to discrete-time circuit models, as we demonstrate below.

We consider a chain with one Majorana operator $\psi_i$ on each site ($i=1,2,\ldots,L$), satisfying $\psi_i=\psi_i^{\dagger}$ and $\{\psi_i,\psi_j\}=2\delta_{ij}$.
At $t=1$, we measure the odd bonds ${\rm i}\psi_{2i-1}\psi_{2i}$ ($i=1,2,\ldots,L/2$) with strength $\Gamma_o=\Gamma(1+\Delta)$.
The corresponding Kraus operators are
\begin{equation}
\label{eq:K_odd}
    \mK_{2i-1,\pm}=(2\cosh \Gamma_o)^{-1} \exp\!\left(\pm \frac{{\rm i}\Gamma_o}{2}\,\psi_{2i-1}\psi_{2i}\right),
\end{equation}
where $\pm$ labels the measurement outcome.
At $t=2$, we measure the even bonds ${\rm i}\psi_{2i}\psi_{2i+1}$ with strength $\Gamma_e=\Gamma(1-\Delta)$, with Kraus operators
\begin{equation}
    \mK_{2i,\pm}=(2\cosh \Gamma_e)^{-1}\exp\!\left(\pm \frac{{\rm i}\Gamma_e}{2}\,\psi_{2i}\psi_{2i+1}\right).
\end{equation}
The state is updated conditioned on the measurement outcomes, which occur with Born probabilities: 
the probability of applying a Kraus operator $\mK$ to a state $\rho$ is $\Tr\,(\mK\rho\,\mK^{\dagger})$.
Repeating the two layers at $t=1$ and $t=2$ generates the measurement-only dynamics.

The single-particle representation of $\mK_{2i-1,+}$ can be written as $K=e^H$, where $H$ equals
$H|_{\{2i-1,2i\}} = {\rm i}\Gamma_o \begin{pmatrix} 0 & 1 \\ -1 & 0 \end{pmatrix}$ \kk{on the bond $\{ 2i-1, 2i\}$} and is zero on all the other sites.
The generator $H$ satisfies particle-hole symmetry $H=-H^{\rm T}$ and chiral symmetry $H=-\Sigma H^{\dagger}\Sigma^{-1}$ with $\Sigma= \sigma_z \otimes 1_{N/2}$, and therefore belongs to class BDI.
The same particle-hole and chiral symmetries 
hold for $\mK_{2i-1,-}$ and for $\mK_{2i,\pm}$, implying that the full measurement-only circuit dynamics is in class BDI.
They exhibit $\mathbb{Z}$-classified topology (see Table~\ref{tab: topology}), characterized by the winding number for $\bar{H}_t$ and the Chern number for $L_t$.

For $\Delta < 0$, the stronger measurements on the even pairs ${\rm i} \psi_{2i} \psi_{2i+1}$ enhance pairing between them, leaving the Majoranas at the edges unpaired under the open boundary conditions, reminiscent of topological superconducting wires~\cite{kitaev01}. 
We numerically simulate the dynamics starting from an initial state characterized by $ {\rm i}\psi_{2i-1}\psi_{2i}=1$ for $i=1,2,\ldots,L/2$ ($\Delta  = \pm 0.8, \Gamma = 1$).
We track the evolution of the correlation matrix 
\begin{equation}
    (D_t)_{ij}\coloneqq \langle \Psi_t|\,{\rm i}[\psi_i,\psi_j]\,|\Psi_t\rangle 
\end{equation}
and extract Lyapunov exponents using a QR-based method (see Appendix~\ref{app: numerical method} for details).
Notably, $D_t$ respects the same particle-hole and chiral symmetries as $H$. 
To reach the steady state, we simulate up to time $T\gg L$.

As discussed in Sec.~\ref{subsec: BBC}, we diagnose the topology of the dynamics from the steady-state correlation matrix $D_T$ via local topological markers.
To this end, we introduce a position operator $X$ that commutes with the chiral operator $\Sigma$, and group sites $2i-1$ and $2i$ into the same unit cell, resulting in $X_{ij}=\lceil i/2\rceil\,\delta_{ij}$.
The local chiral index $\nu_{\mathbb{Z}}(x)$ at unit cell $x$ is~\cite{mondragonShem2014, hannukainen2022}
\begin{equation}
\label{eq: chiral marker}
\nu_{\mathbb{Z}}(x)=\sum_{i=2x-1}^{2x}(D_T\,\Sigma\,X\,D_T)_{ii}.
\end{equation}
We confirm $\nu_{\mathbb{Z}}=1$ ($\nu_{\mathbb{Z}}=0$) for $\Delta<0$ ($\Delta>0$) [Fig.~\ref{fig: chain}\,(c)].

We next examine the Lyapunov spectrum.
Under the periodic boundary conditions, the spectrum is gapped in both the topological and trivial phases (i.e., $\min_n|\eta_n|>0$).
In contrast, under the open boundary conditions, the topological phase exhibits a Lyapunov zero mode [Fig.~\ref{fig: chain}\,(e)].
Notably, the magnitude of this near-zero exponent decreases exponentially with the system size down to numerical precision, indicating a genuine zero mode in the infinite-size limit [see the inset of Fig.~\ref{fig: chain}\,(e)].
Such a zero mode realizes the bulk-boundary correspondence and stabilizes algebraically slow purification dynamics.

The nontrivial topology is also relevant to the universality classes of measurement-induced phase transitions.
In class BDI, 
the perturbative expansion of the beta function for the corresponding nonlinear sigma model reads,
\begin{equation}
\beta \left( t \right) = d-1 - 4t^3 + \mathcal{O}\,( t^4 ), 
\end{equation}
with the coupling parameter $t \geq 0$~\cite{hikami1981, wegner1989, wang2023, evers2008a}.
We always have $\beta < 0$ for $d\leq 1$, precluding phase transitions.
Nevertheless, numerical simulations of lattice models support their presence~\cite{nahum2020a, merritt2023a, fava2023, unpublished}.
This implies that the phase transitions therein cannot be explained by the standard nonlinear sigma model but that with a topological term.
Consistently, the $\mathbb{Z}$-classified topology in Table~\ref{tab: topology} corresponds to a $\theta$ term, inducing critical behavior even for $d=1$, akin to the quantum Hall transitions~\cite{QHE-textbook, Huckestein-review, Kramer-review}.

\subsection{Monitored Majorana fermions in class D}

We supplement the measurement-only circuit with random unitary gates.
At $t=3$, we apply random unitary gates on the odd bonds,
$U_{2i-1} = e^{\theta_{2i-1}\psi_{2i-1}\psi_{2i}/2}$ for $i=1,2,\ldots,L/2$;
at $t=4$, we apply analogous gates on the even bonds,
$U_{2i} = e^{\theta_{2i}\psi_{2i}\psi_{2i+1}/2}$ [see Fig.~\ref{fig: chain}\,(a)].
The real random variables $\{\theta_i\}$ are drawn independently and uniformly from $[-W/2,\,W/2]$, where $W\in[0,2\pi]$ controls the strength of the unitary dynamics.
Repeating the measurements at $t=1,2$ and the unitary layers at $t=3,4$ generates the monitored dynamics.
The single-particle representation of $U_i=e^{\theta_i\psi_i\psi_{i+1}/2}$ can be written as $e^{H^{\prime}}$, where $H^{\prime}$ equals
$H^{\prime}|_{\{i,i+1\}}=\theta_i\begin{pmatrix}0&1\\-1&0\end{pmatrix}$
on the bond $\{i,i+1\}$ and vanishes on all the other sites.
We have $H^{\prime}=-(H^{\prime})^{\rm T}$ but $H^{\prime} \neq -\Sigma (H^{\prime})^{\dag}\Sigma^{-1}$, 
and hence $H^{\prime}$ preserves particle-hole symmetry but breaks chiral symmetry.
Therefore, the resulting generic monitored dynamics belongs to class D, exhibiting the $\mathbb{Z}_2$ classification (Table~\ref{tab: topology}).

We similarly simulate the time evolution via the correlation matrix $D_t$, and take $T\gg L$ to reach the steady state ($\Delta  = \pm 0.8, \Gamma = 1$, and $W = 0.4$).
The steady-state correlation matrix $D_T$ no longer respects chiral symmetry due to the unitary gates generated by $H'$.
To formulate a local $\mathbb{Z}_2$ topological marker~\cite{hannukainen2022}, we define a chiral version of the correlation matrix $Q_T$ by
\begin{equation}
    Q_T = \frac{1}{2}\Big(D_T + {\rm i}\,|[D_T,\Sigma]|^{-1}[D_T,\Sigma]\Big),
\end{equation}
where $\Sigma$ is the chiral symmetry operator of the measurement-only dynamics, and $|\cdot|$ denotes the maximal modulus of an operator.
In our simulations, we find $|[D_T,\Sigma]|>0$.
Replacing $D_T$ by $Q_T$ in Eq.~\eqref{eq: chiral marker}, we define the local $\mathbb{Z}_2$ index as $\nu_{\mathbb{Z}_2}(x)\equiv \nu_{\mathbb{Z}}(x)\,(\mathrm{mod}\ 2)$.
We find $\nu_{\mathbb{Z}_2}=1$ ($\nu_{\mathbb{Z}_2}=0$) for $\Delta<0$ ($\Delta>0$) [Fig.~\ref{fig: chain}\,(d)].
In the topological phase with $\nu_{\mathbb{Z}_2}=1$, a Lyapunov zero mode persists under the open boundary conditions [Fig.~\ref{fig: chain}\,(f)].

The distinction between the $\mathbb{Z}$ and $\mathbb{Z}_2$ topology is illustrated by coupling two topologically nontrivial Majorana chains [Fig.~\ref{fig: chain}\,(b)].
We introduce a symmetry-preserving interchain unitary gate $e^{\phi_i \psi_i^A \psi_i^B}$, where $A$ and $B$ label the two chains and $\phi_i\in[-W^{\prime}/2,\,W^{\prime}/2]$ are independent real random variables that set the interchain coupling strength.
The chiral symmetry operator is $\Sigma\otimes 1$, where $1$ denotes the identity in the $(A,B)$ chain space.
In our simulations, we keep all other parameters unchanged and fix $W^{\prime}=1$.
While the two Lyapunov zero modes survive in class BDI, they are lifted in class D, consistent with the $\mathbb{Z}_2$ classification [Figs.~\ref{fig: chain}\,(e) and (f)].

\begin{figure}[tb]
  \centering
  \includegraphics[width=1\linewidth]{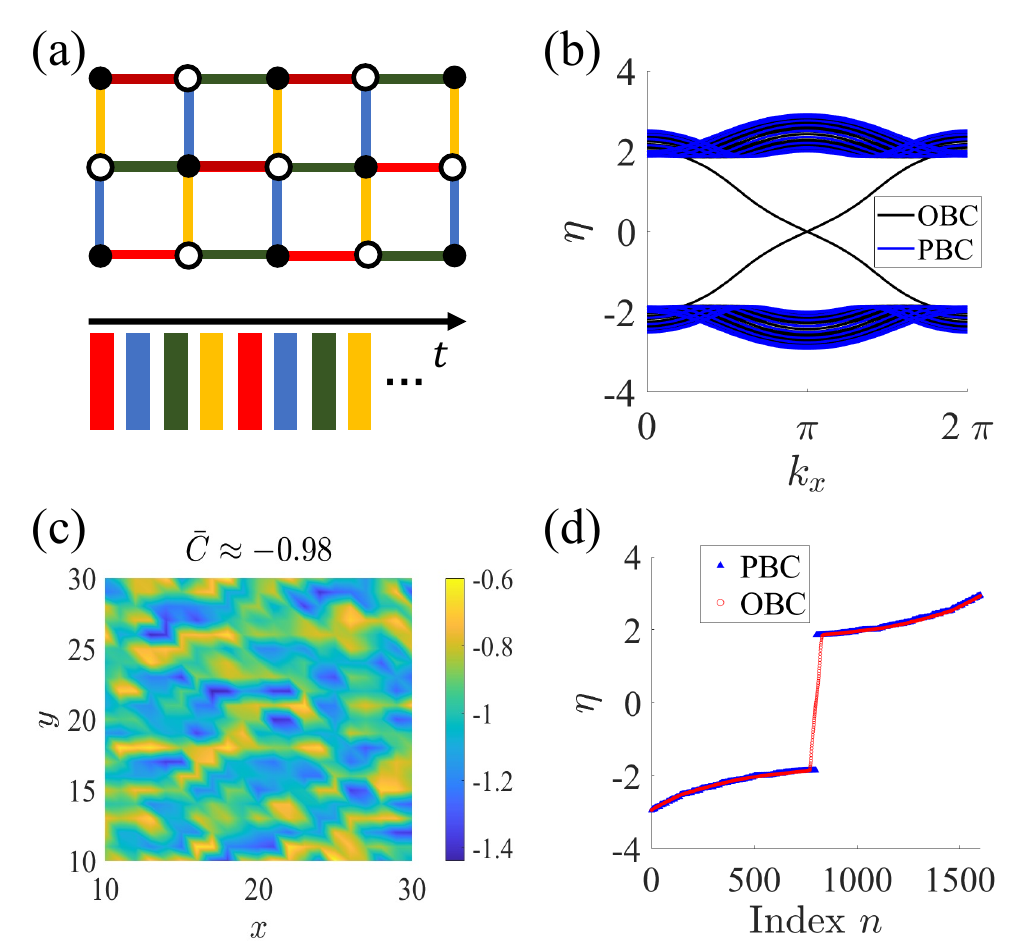}
  \caption{(a)~$\left( 2+1 \right)$-dimensional quantum circuit on a lattice of size $L_x \times L_y$. 
  Each site $\br$ incorporates one fermion $c_{\br}^{\dag}$. 
  At each time step, measurements and unitary gates are applied to the bonds of a specific color, based on the sequence shown at the bottom. 
  (b)~Lyapunov spectra of the dynamics with the homogeneous measurement strength under periodic boundary conditions (PBC) along the $x$ direction, and PBC or open boundary conditions (OBC) along the $y$ direction.
  (c), (d)~Monitored dynamics with inhomogeneous measurement strength.
  (c)~Local Chern marker of the steady state.
  (d)~Lyapunov spectra under PBC and OBC.} 
  \label{fig: Chern}
\end{figure}

\section{Topological monitored dynamics in \texorpdfstring{$2+1$}{2+1} dimensions}
\label{sec: num 2D}

We further investigate monitored complex fermions on a square lattice.
Each site $\br=(x,y)$ hosts a single fermionic mode $c_{\br}^{\dagger}$.
At each time step, unitary gates and measurements are applied to bonds of a given color, following the sequence shown at the bottom of Fig.~\ref{fig: Chern}\,(a).
The unitary gate on the bond connecting sites $\br$ and $\br'$ is
\begin{equation}
U=\exp\!\left[{\rm i}\theta\, t_{\br\br'}\,(c_{\br}^{\dagger}c_{\br'}+c_{\br'}^{\dagger}c_{\br})\right].
\end{equation}
The hopping amplitudes $t_{\br\br'}$ follow the Harper-Hofstadter model~\cite{harper1955} with flux $1/2$ per plaquette:
$t_{\br,\br+{\bm e}_x}=t$ and $t_{\br,\br+{\bm e}_y}=t(-1)^x$.
We weakly measure the occupations of the bonding and antibonding modes, $d^{\dagger}d$ and $f^{\dagger}f$, with
$d\coloneqq (c_{\br}+c_{\br'})/\sqrt{2}$ and $f\coloneqq (c_{\br}-c_{\br'})/\sqrt{2}$.
The corresponding Kraus operators are
\begin{align}
\mK_{d\pm} &=(2\cosh\Gamma)^{-1}e^{\pm\Gamma(d^{\dagger}d-1/2)}, \\
\mK_{f\pm} &=(2\cosh\Gamma)^{-1}e^{\pm\Gamma(f^{\dagger}f-1/2)}. 
\end{align}
We postselect measurement outcomes as follows:
for $t_{\br\br'}=t$, we select $\mK_{d+}$ and $\mK_{f-}$; 
for $t_{\br\br'}=-t$, we select $\mK_{d-}$ and $\mK_{f+}$.
The resulting dynamics has no symmetry and thus belongs to class A, characterized by a three-dimensional winding number for $L_t$ and a Chern number for $\bar{H}_t$.

We first consider the translation-invariant setting in which the measurement strength is uniform on all bonds and time independent, preserving translation symmetry in both space and time.
We simulate the dynamics using the correlation function 
\begin{equation}
(D_t)_{\br\br'}\coloneqq \langle \Psi_t|c_{\br}^{\dagger}c_{\br'}|\Psi_t\rangle
\end{equation}
with $t=1$, $\theta=\pi/3$, and $\Gamma=1$.
We consider a system of size $L\times L$ with $L=40$ and evolve to time $T\gg L$.
We evaluate the topological invariant via the local Chern marker of the steady state.
The local Chern marker $C(\br)$ is determined by the correlation matrix $D_T$ as~\cite{prodan2010, hannukainen2022}
\begin{equation}
    C(\br)=\frac{1}{2\pi{\rm i}}\left(D_T X D_T Y D_T - D_T Y D_T X D_T\right)_{\br,\br},
\end{equation}
where the position operators are $X_{\br,\br'} = x \delta_{\br,\br'}$ and $Y_{\br,\br'} = y \delta_{\br,\br'}$.
We find that $C(\br)$ is quantized to $C(\br)\approx -1$ in the central region of the lattice.
We next impose the periodic boundary conditions along the $x$ direction and the open boundary conditions along the $y$ direction, keeping all the parameters unchanged.
Owing to translation symmetry along the $x$ direction, the cumulative Kraus operator $K_{[0,t]}$ is block diagonal in momentum space, and the Lyapunov exponents can be labeled by the conserved quantum number $k_x$.
The nontrivial Chern number is accompanied by gapless modes in the Lyapunov spectrum [Fig.~\ref{fig: Chern}\,(b)], analogous to chiral edge modes in the quantum Hall effect.

Next, we consider a setting in which the measurement strength fluctuates randomly in both space and time.
Specifically, for each bond at each time step we draw the measurement strength independently and uniformly from $[\Gamma-W/2,\,\Gamma+W/2]$.
We choose $W=0.4$ and keep all the other parameters unchanged.
While the local Chern marker $C(\br)$ becomes spatially inhomogeneous, its average $\bar{C}(\br)$ over the central region remains close to $-1$ [Fig.~\ref{fig: Chern}\,(c)].
Consistently, the Lyapunov spectrum under the open boundary conditions retains gapless modes [Fig.~\ref{fig: Chern}\,(d)].



\section{Discussion}
\label{sec: discussion}

In this work, we establish the tenfold classification of symmetry and topology for monitored free fermions.
Our classification represents an open quantum analog of the periodic table of topological insulators and superconductors~\cite{schnyder08, kitaev09, ryu10, hasan10, qi11, chiu2016b}.
It is independent of the specific measurement protocol and applies both with and without postselection.
In our numerics, we study monitored $(1+1)$-dimensional Majorana fermions without postselection and monitored $(2+1)$-dimensional complex fermions with postselection, thereby illustrating this generality.

We formulate the classification in terms of appropriate gap conditions on the relevant operators.
These gap conditions place the dynamics in a purifying phase, characterized by a finite purification time, playing a role analogous to an insulator phase in the equilibrium classification.
Notably, in most settings, purifying phases having a finite purification time and disentangled phases obeying an area law of entanglement are equivalent, differing primarily in perspective (see, e.g., Ref.~\cite{gullans2020d}).
Extending the topological classification to mixed phases with divergent purification time and to entangled phases with volume-law entanglement is an interesting direction for future work, analogous to the role of gapless phases such as topological semimetals in equilibrium systems~\cite{armitage2018}.

Based on this classification, we elucidate the bulk-boundary correspondence in monitored quantum dynamics, 
which has two complementary implications.
On the one hand, with open spatial boundaries, the spacetime topology of the evolution enforces gapless boundary Lyapunov modes, leading to a topologically protected divergence of the purification time.
On the other hand, the steady state may be viewed as a boundary in the temporal direction and is topological, as diagnosed by local topological markers.
The latter also provides a practical diagnostic of the spacetime topology and is relevant for preparing topological states of matter in quantum-circuit platforms.

Our findings provide a guiding principle to investigate monitored free fermions across various symmetry classes and spacetime dimensions.
For example, a recent work~\cite{pan2024} has found a unique entanglement structure due to the presence of a domain wall.
It merits further study to explore its possible connection with our framework.
Moreover, it should be significant to incorporate many-body interactions into our framework, akin to equilibrium counterparts.
It is also worth noting that non-Hermitian topology causes anomalous boundary phenomena, including the skin effect~\cite{Lee-16, YW-18-SSH, Kunst-18, Yokomizo-19, Zhang-20, OKSS-20}, which should also be relevant to the universality classes of measurement-induced phase transitions~\cite{kawabata2023a, wang2024, lee2024, liu_li_xu2024}.
Finally, we note that neutral atom arrays can realize fermionic quantum processors~\cite{gonzalez-cuadra2023,ott2025} and should be the ideal experimental platforms for our findings.

\smallskip
{\it Note added}.---After the completion of this work, we became aware of a recent related work~\cite{oshima2024}.

\smallskip
\section*{Acknowledgment}
We thank Tomi Ohtsuki, Shinsei Ryu, and Ryuichi Shindou for helpful discussion.
We thank Ryusuke Hamazaki for coordinating the submission of this work and Ref.~\cite{oshima2024}.
Z.X. is supported by the National Basic Research Programs of China (No.~2019YFA0308401) and by the National Natural Science Foundation of China (No.~11674011 and No.~12074008).
Z.X. is supported by the Princeton Quantum Initiative Fellowship.
K.K. is supported by JSPS KAKENHI Grant No.~JP24H00945, No.~JP26H02015, No.~JP26K06970, and No.~JP26K17046.

\appendix
\setcounter{secnumdepth}{2}

\section{Classifying space of \texorpdfstring{$L_t$}{Lt}}
\label{sec: Lt}

As discussed in 
\kk{Sec.~\ref{sec: topology},}
the dynamics in purifying phases requires its non-Hermitian dynamical generator $L_t$ to exhibit a mobility gap at zero.
Since in-gap localized states do not generally influence topology of an operator,
we identify the classifying space of $L_t$ by assuming a spectral gap instead of a mobility gap for simplicity.
We perform the polar decomposition of $L_t$:
\begin{equation}
    L_t = UP, 
\end{equation}
where $U$ is unitary, and $P$ is positive definite owing to $\det L_t \neq 0$.
This decomposition is unique since $P$ is determined as 
$P = (L_t^{\dag} L_t)^{1/2}$.

Importantly,  $U$ shares the same symmetries as $L_t$. 
Substituting $L_t = UP$ to time-reversal symmetry in Eq.~(\ref{eq: TRS-L}),
we have 
\begin{equation} 
  L_t  = (\mathcal{T} U^* \mathcal{T}^{-1}) (\mathcal{T} P^* \mathcal{T}^{-1}).
\end{equation} 
The uniqueness of the polar decomposition requires that $U$ also respects time-reversal symmetry: 
$\mathcal{T} U^* \mathcal{T}^{-1} = U$.
Particle-hole symmetry in Eq.~(\ref{eq: PHS-L}) 
leads to
\begin{equation} 
     L_t =  (-\mC U^{\rm T} \mC^{-1}) (\mC U^* P^{\rm T} U^{\rm T} \mC^{-1}),
\end{equation}
and thus
$\mC U^{\rm T} \mC^{-1} = -U$.
Similarly, chiral symmetry of $L_t$ in Eq.~(\ref{eq: CS-L}) results in $\Gamma U^{\dagger} \Gamma^{-1} = - U$.

Furthermore, $L_t$ can be continuously deformed into $U$ through the path: 
\begin{equation}
U \left[ (1 - \lambda) P + \lambda I \right] \quad \left( 0 \leq \lambda \leq 1 \right)
\end{equation}
with the identity matrix $I$. 
Since $P$ is positive definite, we have $\det\left[ (1 - \lambda) P + \lambda I \right] \neq 0$ for arbitrary $\lambda \in [0,1]$, ensuring that the gap remains open during this deformation.
In addition, symmetry of $L_t$ is preserved for arbitrary $\lambda \in [0,1]$.
Therefore, $U$ and $L_t$ share the same classifying space. 

We identify the classifying space of $L_t$ by the associated unitary operator $U$, as follows.

{\it Classes A, AI, and AII}.---Classes A, AI, and AII are concerned solely with time-reversal symmetry in Eq.~(\ref{eq: TRS-L}) (Table~\ref{tab: symmetry}).
No symmetry is preserved in class A, while time-reversal symmetry with $\mathcal{T}\mathcal{T}^{*} = +1$ ($-1$) is respected in class AI (AII).
Depending on the presence or absence of time-reversal symmetry, 
$U$ associated with $N \times N$ ($2N \times 2N$) non-Hermitian dynamical generators $L_t$ in classes A and AI (class AII) generally belongs to
\begin{equation}
    U \in \begin{cases}
        \mathrm{U} \left( N \right) \cong \mathcal{C}_1 & \left( \text{class~A} \right); \\
        \mathrm{O} \left( N \right) \cong \mathcal{R}_1 & \left( \text{class~AI} \right); \\
        \mathrm{Sp} \left( N \right) \cong \mathcal{R}_5 & \left( \text{class~AII} \right), \\
    \end{cases}
\end{equation}
which directly follows from the definitions of the unitary group $\mathrm{U} \left( N \right)$, orthogonal group $\mathrm{O} \left( N \right)$, and symplectic group $\mathrm{Sp} \left( N \right)$.

{\it Classes AIII, CI, and DIII}.---Chiral symmetry ($ \Gamma U^{\dagger} \Gamma^{-1} = -U  $) is common among classes AIII, CI, and DIII, and time-reversal symmetry ($ \mathcal{T} U^* \mathcal{T}^{-1} = U  $) with $\mathcal{T}\mathcal{T}^{*} = +1$ ($-1$) is additionally respected in class CI (DIII).
Chiral symmetry implies Hermiticity $\left( \ii U \Gamma \right)^{\dag} = \ii U \Gamma$ and $\left( \ii U \Gamma \right)^2 = 1$.
Time-reversal symmetry imposes $\mathcal{T} \left( \ii U \Gamma\right)^{*} \mathcal{T}^{-1} = \ii U \Gamma$.
Then, $\ii U \Gamma$ can be generally diagonalized via a unitary matrix $V$:
\begin{equation}
    U = - \ii V \begin{pmatrix}
        1_M & 0 \\
        0 & - 1_N
    \end{pmatrix} V^{-1} \Gamma,
\end{equation}
with the $M \times M$ ($N \times N$) identity matrix $1_M$ ($1_N$) and
\begin{equation} 
    V \in \begin{cases}
        \mathrm{U} \left( M+N \right) & \left( \text{class~AIII} \right); \\
        \mathrm{O} \left( M+N \right) & \left( \text{class~CI} \right); \\
        \mathrm{Sp} \left( M+N \right) & \left( \text{class~DIII} \right). \\
    \end{cases}
\end{equation}
Additionally, this diagonalization is invariant under the gauge transformation 
    \begin{equation} \label{eq: gauge1}
        V \mapsto V \begin{pmatrix}
            \tilde{V}_M & 0 \\
            0 & \tilde{V}_N
        \end{pmatrix},~\tilde{V}_N \in \begin{cases}
            \mathrm{U} \left( N \right) & \left( \text{class~AIII} \right); \\
            \mathrm{O} \left( N \right) & \left( \text{class~CI} \right); \\
            \mathrm{Sp} \left( N \right) & \left( \text{class~DIII} \right). \\
        \end{cases}
    \end{equation}
Consequently, the classifying spaces are respectively given as the complex, real, and quaternionic Grassmannians:
\begin{equation}
    V \in \begin{cases} 
        \mathrm{U} \left( M+N \right)/\mathrm{U} \left( M \right) \times \mathrm{U} \left( N \right) \cong \mathcal{C}_0 & \left( \text{class~AIII} \right); \\
        \mathrm{O} \left( M+N \right)/\mathrm{O} \left( M \right) \times \mathrm{O} \left( N \right) \cong \mathcal{R}_0 & \left( \text{class~CI} \right); \\
        \mathrm{Sp} \left( M+N \right)/\mathrm{Sp} \left( M \right) \times \mathrm{Sp} \left( N \right) \cong \mathcal{R}_4 & \left( \text{class~DIII} \right). \\
        \end{cases}
\end{equation}

{\it Classes BDI and CII}.---In class BDI, as a result of time-reversal symmetry, the unitary Hermitian matrix $\ii U \Gamma$ is subject to particle-hole symmetry $\mathcal{T} \left( \ii U \Gamma \right)^{*} \mathcal{T}^{-1} = - \ii U \Gamma$.
Here, let us choose $\mathcal{T}$ as the $2N \times 2N$ identity matrix $1_{2N}$ without loss of generality.
Then, since $U \Gamma$ is a real antisymmetric matrix, it can be diagonalized in a proper basis as
\begin{equation}
    U = O
    \left( \ii \Sigma_y \right) O^{-1} \Gamma,
\end{equation}
with $O \in \mathrm{O} \left( 2N \right)$ and $\Sigma_y \coloneqq \sigma_y \otimes 1_N$.
This orthogonal matrix $O$ obeys the gauge transformation $O \mapsto O\tilde{O}$ satisfying
\begin{equation}
    \tilde{O} 
    \Sigma_y \tilde{O}^{-1} = 
    \Sigma_y, \quad \tilde{O} \in \mathrm{O} \left( 2N \right).
\end{equation}
When we introduce a unitary matrix $G \coloneqq \left( 1_{2N} + \Sigma_y \right)/\sqrt{2}$ that transforms $\Sigma_z \coloneqq \sigma_z \otimes 1_N$ to $\Sigma_y$ (i.e., $G \Sigma_z G^{-1} = \Sigma_y$), 
the above gauge transformation reduces to
\begin{equation}
    ( G^{-1} \tilde{O} G ) 
    \,\Sigma_z\,
    ( G^{-1} \tilde{O} G )^{-1} = 
    \Sigma_z.
\end{equation}
Hence, the allowed gauge transformation is
\begin{equation}
    \tilde{O} = G \begin{pmatrix}
        W & 0 \\
        0 & W^{*}
    \end{pmatrix} G^{-1}, \quad W \in \mathrm{U} \left( N \right).
\end{equation}
Accordingly, the classifying space is 
\begin{equation}
    O \in \mathrm{O} \left( 2N \right)/\mathrm{U} \left( N \right) \cong \mathcal{R}_2 \quad \left( \text{class BDI} \right).
\end{equation}

In class CII, let us choose $\mathcal{T}$ for time-reversal symmetry as $\mathcal{T} = \Sigma_y$.
Owing to time-reversal symmetry, $U$ can be diagonalized in a proper basis as
\begin{equation}
    U = V 
    \left( \ii \Sigma_y \right) V^{-1} \Gamma,
\end{equation}
with $V \in \mathrm{Sp} \left( N \right)$.
Since $V$ has gauge ambiguity in a similar manner to class BDI, the classifying space is 
\begin{equation}
    V \in \mathrm{Sp} \left( N \right)/\mathrm{U} \left( N \right) \cong \mathcal{R}_6 \quad \left( \text{class CII} \right).
\end{equation}

{\it Classes D and C}.---In class D, $U$ respects particle-hole symmetry: $\mathcal{C} U^{\rm T} \mathcal{C}^{-1} = -U $ with $\mathcal{C}\mathcal{C}^{*} = +1$, where $\mathcal{C}$ is chosen as $\mathcal{C} = 1_{2N}$ without loss of generality.
Then, particle-hole symmetry reduces to $\Sigma_y \left( \Sigma_y U \right)^{\rm T} \Sigma_y^{-1} = \Sigma_y U$.
Hence, $U$ is generally expressed as
\begin{equation}
    U = \Sigma_y f^{\rm T} \Sigma_y f \Sigma_y, \quad f \in \mathrm{U} \left( 2N \right),
\end{equation}
and has the gauge ambiguity $f \mapsto gf$ with $g \in \mathrm{Sp} \left( N \right)$.
Consequently, the classifying space is
\begin{equation}
    f \in \mathrm{U} \left( 2N \right)/\mathrm{Sp} \left( N \right) \cong \mathcal{R}_3 \quad \left( \text{class D} \right).
\end{equation}

In class C, $L$ respects particle-hole symmetry: $\mathcal{C} U^{\rm T} \mathcal{C}^{-1} = -U $ with $\mathcal{C}\mathcal{C}^{*} = -1$, where $\mathcal{C}$ is chosen as $\mathcal{C} = \Sigma_y$ without loss of generality.
Then, particle-hole symmetry reduces to $\left( \Sigma_y U \right)^{\rm T} = \Sigma_y U$.
Hence, $U$ is generally expressed as
\begin{equation}
    U = \Sigma_y f^{\rm T} f, \quad f \in \mathrm{U} \left( N \right),
\end{equation}
and has the gauge ambiguity $f \mapsto gf$ with $g \in \mathrm{O} \left( N \right)$.
Consequently, the classifying space is
\begin{equation}
    f \in \mathrm{U} \left( N \right)/\mathrm{O} \left( N \right) \cong \mathcal{R}_7 \quad \left( \text{class C} \right).
\end{equation}

The topological classification can also be understood through a Hermitized operator~\cite{gong2018, kawabata2019}
\begin{equation}
    \tilde{L}_t \coloneqq \begin{pmatrix}
        0 & L_t \\
        L_t^{\dag} & 0
    \end{pmatrix}.
\end{equation}
If $L_t$ has a mobility gap for $z = 0$, Hermitian $\tilde{L}_t$ also has a mobility gap at its spectral origin, and vice versa.
Therefore, $L_t$ and $\tilde{L}_t$ share the same classifying space and topological classification.
By construction, $\tilde{L}_t$ respects additional chiral symmetry, $\Sigma_z \tilde{L}_t \Sigma_z^{-1} = - \tilde{L}_t$ with a Pauli matrix $\Sigma_z$, changing the relevant symmetry classes, as also listed in Table~\ref{tab: symmetry}.

\section{Classifying space of \texorpdfstring{$\bar{H}_t$}{barHt}}
\label{sec: Ht}

Finite purification time requires $\bar{H}_t$ to possess a gap with respect to ${\rm Re}\,z = 0$.
Below, we demonstrate that $\bar{H}_t$ with this real line gap condition can be continuously deformed into a Hermitian one while preserving the gap and symmetry.
Although the following derivation is based on the Schur decomposition and hence different from the approach in Ref.~\cite{kawabata2019}, we reach the same conclusion.

Let us perform the Schur decomposition of $\bar{H}_t$: 
$\bar{H}_t = Q r Q^{\dag}$ with a unitary matrix $Q$ and an upper triangular matrix $r$.
The diagonal elements of $r$ coincide with eigenvalues of $\bar{H}_t$ (i.e., $r_{ii} = z_i$). 
We order $r_{ii}$'s by ${\rm Re}\,r_{ii} > 0$ ($1 \leq i \leq p$) and ${\rm Re}\,r_{ii} < 0$ ($p+1 \leq i \leq N-p$), where $p$ and $N-p$ denote the numbers of $r_{ii}$'s with positive and negative real parts, respectively.
For this ordering of eigenvalues, the Schur decomposition is unique up to a gauge transformation.  
Let $\bar{H}_t = Q^{\prime}r^{\prime}(Q^{\prime})^\dagger$ be an alternative Schur decomposition such that $\mathrm{Re}\,r^{\prime}_{ii} > 0$ for $1 \leq i \leq p$ and $\mathrm{Re}\,r^{\prime}_{ii} < 0$ for $p+1 \leq i \leq N$.  
We then have $Q^{\prime}= Q \tU$ and $r^{\prime}= \tU^{-1} r \tU$, where $\tU$ is a block-diagonal unitary matrix with blocks of size $p \times p$ and $(N-p) \times (N-p)$ [i.e., $\tU \in \mathrm{U}(p) \times \mathrm{U}(N-p)$].
If $z_i$'s are nondegenerate and we have $r_{ii} = r^{\prime}_{ii}$ for all $i \leq N$, $\tU$ can be further restricted to $\mathrm{U} (1)^{N}$.
However, $\tU \in \mathrm{U}(p) \times \mathrm{U}(N-p)$ suffices for the subsequent discussion.

We show that symmetry of $\bar{H}_t$ also leads to corresponding symmetry of $Q$.
Applying time-reversal symmetry in Eq.~(\ref{eq: TRS-L}) to $\bar{H}_t = Q r Q^{\dag}$, we have 
\begin{equation} 
\bar{H}_t = \mathcal{T} Q^* r^* Q^{\rm T} \mathcal{T}^{-1} ,
\end{equation}
which represents an alternative Schur decomposition and satisfies ${\rm Re}\,r_{ii} = {\rm Re}\,r^*_{ii}$.
The uniqueness of the Schur decomposition leads to $\mathcal{T} Q^* = Q \tU$ with $\tU \in \mathrm{U}(p) \times \mathrm{U}(N-p)$.
Particle-hole symmetry in Eq.~(\ref{eq: PHS-L}) gives
\begin{equation} \label{eq: Schur PHS}
\bar{H}_t = -\mathcal{C} Q^* r^{\rm T} Q^{\rm T} \mathcal{C}^{-1}.
\end{equation}
Particle-hole symmetry makes the eigenvalues appear in $(z_i, -z_i)$ pairs and enforces $p = N-p$, further indicating that $\{r_{ii}\,|\,1 \leq i \leq p \}$ is identical to $\{-r_{ii}\,|\,p+1 \leq i \leq N \}$. 
Although $r^{\rm T}$ is not upper triangular, $V r^{\rm T} V^{-1}$ is upper triangular for a unitary matrix $V$ defined by $V_{ij} \coloneqq \delta_{i, N+1 -j}$. 
Using $V$ and Eq.~(\ref{eq: Schur PHS}), we have an alternative Schur decomposition:
\begin{equation}
    \bar{H}_t = \mathcal{C} Q^* V^{-1} (-Vr^{\rm T} V^{-1}) V  Q^{\rm T} \mathcal{C}^{-1},
\end{equation}
satisfying $\mathrm{Re}\,(-Vr^{\rm T}V^{-1})_{ii} > 0$ for $1 \leq i \leq p$ and $\mathrm{Re}\,(-Vr^{\rm T}V^{-1})_{ii} < 0$ for $p+1 \leq i \leq N$.
Therefore, we have $\mathcal{C} Q^* V^{-1} = Q   \tU$ with $\tU \in  \mathrm{U}(p) \times \mathrm{U}(N-p)$.
Similarly, chiral symmetry in Eq.~(\ref{eq: CS-L}) leads to $\Gamma Q V^{-1} = Q \tU $ with $\tU \in  \mathrm{U}(p) \times \mathrm{U}(N-p)$.

Next, we introduce a Hermitian Hamiltonian $Q \mE Q^{-1}$ with $\mE \coloneqq {\rm diag}(1_{N},-1_{M})$ and show that it inherits the same symmetry as $\bar{H}_t$.
Time-reversal symmetry of $Q$ (i.e., $\mathcal{T} Q^* = Q \tU$) ensures 
\begin{equation}
    \mathcal{T}(Q \mE Q^{-1})^* \mathcal{T}^{-1}   
    =  Q \mE Q^{-1} ,
\end{equation}
signifying time-reversal invariance also for $Q \mE Q^{-1}$.
Particle-hole symmetry (i.e., $\mathcal{C} Q^* V^{-1} = Q \tU$) and $V \mE V^{-1} = - \mE $ yield 
\begin{equation}
    \mathcal{C}(Q \mE Q^{-1})^{\rm T} \mathcal{C}^{-1}  
    = - Q \mE Q^{-1}.
\end{equation}
Similarly, chiral symmetry of $Q$ leads to 
\begin{equation}
   \Gamma (Q \mE Q^{-1})^{\dagger} \Gamma^{-1} = -Q \mE Q^{-1}.
\end{equation}
The Hamiltonian $\bar{H}_t = Q r Q^{\dagger}$ can be continuously deformed into $Q \mE Q^{-1}$ 
through the path $\bar{H}_t = Q [\lambda \mE + (1 - \lambda) r ] Q^{\dagger}$ with $ \lambda \in [ 0,1]$.
Due to the gap of $\bar{H}_t$ at ${\rm Re}\,z = 0$, any Hamiltonian in this path retains the gap.
Thus, $\bar{H}_t$ and $Q\mE
 Q^{-1}$ belong to the same classifying space, 
which is obtained, e.g., in Ref.~\cite{chiu2016b} (Table~\ref{tab: symmetry}).

We identify the classifying spaces of $\bar{H}_t$ by those of $Q \mathbb{E} Q^{-1}$. In class A, $Q \mathbb{E} Q^{-1}$ satisfies no symmetry except Hermiticity and hence is a complex Hermitian matrix. 
In class AI (AII), $Q \mathbb{E} Q^{-1}$ satisfies time-reversal symmetry with sign $+1$ ($-1$) and hence is a real (quaternionic) Hermitian matrix. 
Thus, for classes A, AI, and AII, $Q$ belongs to the groups $\mathrm{U}(N)$, $\mathrm{O}(N)$, and $\mathrm{Sp}(N)$, respectively.
In addition, $Q$ has gauge ambiguity similar to that in Eq.~(\ref{eq: gauge1}).
Thus, we have 
\begin{equation}
    Q \in \begin{cases} 
        \mathrm{U} \left( N \right)/\mathrm{U} \left( p \right) \times \mathrm{U} \left( N - p \right) \cong \mathcal{C}_0 & \left( \text{class~A} \right); \\
        \mathrm{O} \left( N \right)/\mathrm{O} \left( p \right) \times \mathrm{O} \left( N -p \right) \cong \mathcal{R}_0 & \left( \text{class~AI} \right); \\
        \mathrm{Sp} \left( N \right)/\mathrm{Sp} \left( p \right) \times \mathrm{Sp} \left( N - p \right) \cong \mathcal{R}_4 & \left( \text{class~AII} \right). \\
        \end{cases}
\end{equation}
As mentioned in 
\kk{Sec.~\ref{sec: topology},}
although $\bar{H}_t$ and $L_t$ share the same symmetry, they belong to different classifying spaces, because the former is concerned with a real line gap for ${\rm Re}\,z = 0$ while the latter is only concerned with a point gap for $z=0$. 
For $L_t$ in classes A, AI, and AII, the classifying spaces are $\mathcal{C}_1$, $\mathcal{R}_1$, and $\mathcal{R}_5$, respectively (see Table~\ref{tab: symmetry} and Appendix~\ref{sec: Lt}).

For the remaining symmetry classes, we can identify the classifying spaces similarly, which are summarized in Table~\ref{tab: symmetry}. 
As exemplified by classes A, AI, and AII, we find that if the classifying space of $L_t$ is $\mathcal{C}_s$ ($\mathcal{R}_s$), that of $\bar{H}_t$ is $\mathcal{C}_{s-1}$ ($\mathcal{R}_{s-1}$).

\newcommand{\mQ}{\mathcal{Q}}
\newcommand{\bk}{\bm k}

\section{Topological invariants of \texorpdfstring{$L_t$}{Lt} and \texorpdfstring{$\bar{H}_t$}{barHt}}
    \label{sec: Green}

As discussed in 
\kk{Sec.~\ref{sec: topology},}
$L_t \coloneqq \partial_t - H_t$ is a non-Hermitian operator acting on $(d+1)$-dimensional spacetime, while $\bar{H}_t$, defined by $K_{[0,t]} \eqqcolon e^{\bar{H}_t t}$, represents the time average of $H_t$ and acts on $d$-dimensional space. 
We assume that the temporal fluctuations of $\bar{H}_t$ are negligible for $t \to \infty$, leading to $\bar{H}_t = \bar{H}$, independent of time. 
In the subsequent discussion on topological invariants, we further assume (spatial) translation invariance of $\bar{H}$ for convenience, and its Bloch Hamiltonian is denoted by $\bar{H}(\bm{k})$. 
In this case, the Fourier transform of $L_t$ is given by $L({\omega, \bm{k}}) = {\rm i} \omega - \bar{H}({\bm{k}})$, equivalent to the inverse of the Green’s function, $G^{-1}({\rm i} \omega, \bm{k}) \coloneqq \ii \omega - \bar{H}({\bm{k}})$.

\subsection{Class A}

In class A and $2n$ spatial dimensions ($n \in \mathbb{Z}$), $\bar{H}(\bm{k})$ is characterized by the $n$th Chern number $C_n$ in the presence of a real line gap with respect to $\mathrm{Re}\,z = 0$. 
Meanwhile, $L({\omega, \bm{k}})$ belongs to class A and acts on $2n+1$ dimensions, whose point-gap topology is characterized by the $\left( 2n+1 \right)$-dimensional winding number $W_{2n+1}$.
Both invariants coincide with each other, given as~\cite{ryu10, kawabata2019}
\begin{equation}
    C_n = W_{2n+1} = \frac{n!}{(2\pi {\rm i})^{n+1}(2n+1)!} \int_{(\omega, {\bm{k}})} \mathrm{Tr}\left(L dL^{-1}\right)^{2n+1},
\end{equation}
where $\int_{(\omega, {\bm{k}})}$ denotes the integral over $\left( 2n+1 \right)$-dimensional momentum-frequency space.
This represents the correspondence of topology between $\bar{H}$ and $L_t$ in class A.

\subsection{Class AIII}

In class AIII and $2n-1$ spatial dimensions ($n \in \mathbb{Z}$), $\bar{H}(\bm{k})$ is characterized by the $\left( 2n-1 \right)$-dimensional winding number $W_{2n-1}$ in the presence of a real line gap with respect to $\mathrm{Re}\,z = 0$. 
Without loss of generality, let the chiral operator for $\bar{H}$ be a Pauli matrix $\sigma_z$, i.e., $\sigma_z H^{\dag}(\bm k) \sigma_z = -H (\bm k)$.
Due to the real line gap, 
$H(\bm k)$ can be continuously deformed into a flat-band Hermitian Hamiltonian $h({\bm k})$ in class AIII: 
\begin{equation}
h(\bm{k}) = P(\bk) \begin{pmatrix}
    1_{p\times p} & 0 \\
    0 & -1_{p\times p} 
\end{pmatrix} P^{\dag}(\bk),
\end{equation}
where $2p$ is the number of the bands, and $P$ is a unitary matrix ($PP^{\dag} = P^{\dag}P = 1$). 
Due to chiral symmetry, $P(\bk)$ takes the form of
\begin{equation}
    P(\bk) = \frac{1}{\sqrt{2}} \begin{pmatrix}
    U(\bk) & U(\bk) \\
    V(\bk) & -V(\bk) 
\end{pmatrix},
\end{equation}
with unitary matrices $U$ and $V$,
leading to 
\begin{equation}
    h(\bm{k}) = \begin{pmatrix}
    0 & U(\bk)V^{\dag}(\bk) \\
    V(\bk)U^{\dag}(\bk) & 0 
\end{pmatrix} \, .
\end{equation}
Introducing $Q(\bk) \coloneqq U(\bk)V^{\dag}(\bk)$, we have the $\left( 2n-1 \right)$-dimensional winding number
\begin{equation} \label{eq: W2n-1}
    W_{2n-1} = \frac{(n-1)!}{(2\pi {\rm i})^{n}(2n-1)!} \int_{ \bk} \mathrm{Tr}\left(Q dQ^{-1}\right)^{2n-1} \, ,
\end{equation}
where $\int_{ \bk}$ denotes the integral over $\left( 2n-1 \right)$-dimensional momentum space.

On the other hand, $L({\omega, \bm{k}})$ belongs to class AIII and acts on $2n$ dimensions, whose point-gap topology is characterized by the $n$th Chern number $C_n$, as explained below. 
We introduce a Hermitized operator
\begin{equation}
    \tilde{L}(\omega, \bm{k}) \coloneqq \begin{pmatrix}
        0 & L({\omega, \bm{k}}) \\
       L^{\dag}({\omega, \bm{k}}) & 0
    \end{pmatrix},
\end{equation}
sharing the same topological classification with $L({\omega, \bm{k}})$. 
Owing to chiral symmetry of $H (\bm k)$, $L({\omega, \bm{k}}) = {\rm i} \omega - \bar{H}({\bm{k}})$ also respects chiral symmetry $\sigma_z L^{\dag}({\omega, \bm{k}}) \sigma_z = - L({\omega, \bm{k}})$.
Consequently, after a unitary transformation, $\tilde{L}(\omega, \bm{k})$ can be block diagonalized into 
\begin{equation}
    \begin{pmatrix}
      \tilde{l}(\omega,  {\bm k}) & 0 \\
       0 &  -\tilde{l}(\omega, {\bm k})
    \end{pmatrix}
\end{equation}
with   
\begin{equation}
    \tilde{l}(\omega,  {\bm k}) \coloneqq - \ii L({\omega, \bm{k}}) \sigma_z = \omega \sigma_z + \ii \bar{H}({\bm{k}}) \sigma_z.
\end{equation}
Then, the topological classification of $L(\omega, \bm{k})$ reduces to that of the Hermitian matrix $\tilde{l}(\omega,  {\bm k})$ in class A, classified by the $n$th Chern number $C_n$.
Through the continuous deformation from $\bar{H}(\bk)$ to $h(\bk)$, the energy gap remains open, and we can continuously deform $\tilde{l}(\omega, \bk)$ to $l(\omega, \bk) \coloneqq \omega \sigma_z + \ii h(\bk) \sigma_z$. 
Therefore, $\tilde{l}(\omega, \bk)$ and $l(\omega, \bk)$ share the same Chern number.
In addition, $l(\omega, \bk)$ is diagonalized as 
\begin{equation}
    l(\omega, \bk) = 
     \mU \begin{pmatrix}
        \sqrt{1 + \omega^2}& 0 \\
        0 & -\sqrt{1 + \omega^2} \\
    \end{pmatrix} \mU^{\dag}
\end{equation}
with
\begin{equation}
    \mU \coloneqq
    \frac{1}{\sqrt{1+y^2}} \begin{pmatrix}
        U & \ii y U \\
        \ii y V& V \\
    \end{pmatrix}, \quad y \coloneqq \sqrt{1 + \omega^2} - \omega .
\end{equation}
Furthermore, $l$ is continuously deformable into the flat-band Hamiltonian $\mQ$,
\begin{equation}
    \mQ \coloneqq \mU \begin{pmatrix}
        1 & 0 \\
        0  & -1  \\
    \end{pmatrix} \mU^{\dag} = 
    \frac{1}{\sqrt{ 1+\omega^2 }}
    \begin{pmatrix}
     \omega & -\ii Q \\
       \ii Q^{\dag} & -\omega \\
    \end{pmatrix} \, .
\end{equation}
Then, the $n$th Chern number $C_n$ is given as
\begin{equation}
    C_n =  -\frac{1}{2^{2n+1} n!} \left( \frac{\rm i}{2\pi} \right)^n \int_{(\omega, \bk )} \mathrm{Tr} \left[ \mQ \left(d \mQ \right)^{2n} \right].
\end{equation}

Notably, we have $C_n = -W_{2n-1}$.
As an illustration, let us consider $n=1$.
Using $k_0 \coloneqq \omega$, $\bk \eqqcolon (k_1, k_2, \ldots, k_{2n-1})$, and $\partial_i \coloneqq \frac{\partial}{\partial k_i}$ ($i = 0,1,2,\ldots, 2n-1$), we have
\begin{align}
   \partial_0 \mQ  &= \frac{1}{\left(1+\omega ^2\right)^{3/2}} \left(
\begin{array}{cc}
 1  & \ii \omega Q \\
 -\ii \omega Q^{\dag} & -1 \\
\end{array}
\right) , \\
 \partial_i \mQ  &= \frac{1}{\sqrt{1+\omega^2}}  \begin{pmatrix}
        0 &  -\ii \partial_{i}Q  \\
        \ii \partial_{i}Q^{\dag}  & 0 
    \end{pmatrix} \quad (i \neq 0). 
\end{align}
Then, we have for $n=1$
\begin{align}
    C_1 &= -\frac{\ii}{16 \pi} \int  \Tr( \mQ \partial_0 \mQ \partial_1 \mQ - \mQ \partial_1 \mQ \partial_0 \mQ) d \omega dk_1 \nonumber \\
    &=  \frac{\ii}{16 \pi} \int \frac{4}{(1 + \omega^2)^{3/2}} d \omega 
        \int \Tr (Q \partial_1 Q^{-1}) dk_1 \nonumber \\
    &=  -\frac{1}{2 \pi \ii} \int \Tr (Q \partial_1 Q^{-1}) dk_1 \, ,
\end{align}
identical to $-W_1$ in Eq.~(\ref{eq: W2n-1}).

\section{Monitored complex fermions in \texorpdfstring{$0+1$}{0+1} dimension}
    \label{sec: 0D}

We generally formulate the $\mathbb{Z}$ topological invariant for non-Hermitian dynamical generators $L_t$ in $0 + 1$ spacetime dimension.
We introduce a $\mathrm{U} \left( 1 \right)$ scalar potential $\mu$ and assume that $L_t = L_t \left( \mu \right)$ remains invertible for arbitrary $\mu$, i.e.,
\begin{equation}
    \forall \mu \in \mathbb{R}~~~ \det L_t \left( \mu \right) \neq 0.
\end{equation}
This condition corresponds to the presence of a point gap for non-Hermitian operators $L_t$~\cite{gong2018, kawabata2019}.
Then, the winding number $W_1$ of the determinant of $L_t$ in the complex plane 
can be introduced, giving
the $\mathbb{Z}$ topological invariant for $d+1 = 1$ and class A in Table~\ref{tab: topology}.

While $L_t$ is inherently stochastic, 
let us suppose that $L_t$ is invariant under time translation.
Under this condition, we can perform its Fourier transform $\tilde{L} \left( \omega \right)$, leading to 
\begin{equation}
    W_1 \coloneqq - \oint_{-\infty}^{\infty} \frac{d\omega}{2\pi\ii} \frac{d}{d\omega} \log \det \tilde{L} \left( \omega \right).
\end{equation}
For example, for $L_t = \partial_t - \gamma$, its Fourier transform becomes $\tilde{L} \left( \omega \right) = \ii \omega - \gamma$, yielding $W_1 = \mathrm{sgn} \left( \gamma \right)/2$.
Notably, this winding number $W_1$ also coincides with the zeroth Chern number of $H_t = \partial_t - L_t = \gamma$, consistent with Appendix~\ref{sec: Green}.

To clarify nontrivial topology in zero spatial dimension $d=0$, we consider the random nonunitary quantum dynamics of $N$ complex fermions whose particle numbers $n_i$'s ($i=1, \cdots, N$) are continuously measured.
The associated non-Hermitian dynamical generator $L_t$ is expressed as~\cite{jacobs2006, wiseman2009quantum, zhenyuxiao2024}
\begin{equation}
    L_t = \partial_t + \ii h_t - \gamma \left( 2 \braket{n}_t - 1 \right) - \sqrt{\gamma} w_t,
\end{equation}
where $h_t$ denotes a random time-dependent $N \times N$ Hermitian Hamiltonian, $\gamma > 0$ the measurement strength, $\braket{n}_t \coloneqq \mathrm{diag} \left( \braket{n_1}_t, \cdots, \braket{n_N}_t \right)$ the average particle numbers, and $w_t$ a stochastic noise term.
Since $L_t$ respects no internal symmetry, it falls into class A and exhibits $\mathbb{Z}$ topology (see Tables~\ref{tab: symmetry} and \ref{tab: topology}).
As discussed above, the corresponding $\mathbb{Z}$ topological invariant is the winding number $W_1$ of the complex spectrum of $L_t$ during the adiabatic insertion of the $\mathrm{U} \left( 1 \right)$ scalar potential, reducing to $W_1 = \sum_{i=1}^{N} \mathrm{sgn} \left( \braket{n_i} - 1/2 \right)/2$ on average.

As shown in Ref.~\cite{zhenyuxiao2024}, Born measurements inherently drive the average particle numbers to $\braket{n_i}_t = 0$ or $\braket{n_i}_t = 1$ for $t \to \infty$, yielding a topological mass term accompanied by the exponential decay of entropy.
By contrast, forced measurements can stabilize a special mode with $\braket{n_i}_t = 1/2$, leading to critical behavior with a divergent purification time.
We find that this quantum criticality is protected by the $\mathbb{Z}$ topology, akin to disordered quantum wires with chiral symmetry~\cite{dyson1953, brouwer1998, mondragonShem2014, altland2014a}.
Notably, it also shares the same universality class as the Anderson transitions induced by point-gap topology in nonreciprocal disordered systems;
upon replacing time with space, $L_t$ becomes a continuum counterpart of the Hatano-Nelson model~\cite{Hatano-Nelson-96, Hatano-Nelson-97}.

\section{Numerical method}
\label{app: numerical method}

We describe the numerical algorithm used to simulate monitored Majorana fermions.
The initial Gaussian state $|\Psi_0\rangle$ is specified by a set of annihilation operators
$c_n=\frac{1}{\sqrt{2}}\sum_{i=1}^{L}\psi_i\,U_{in}$ ($n=1,2,\ldots,L/2$),
with
$U=1_{L/2}\otimes \frac{1}{\sqrt{2}}\begin{pmatrix}1\\ -\ii\end{pmatrix}$.

We consider a weak measurement of ${\rm i}\psi_{2i-1}\psi_{2i}$, described by the Kraus operator
$\mK_{2i-1,\pm}=(2\cosh \Gamma_o)^{-1}\exp\!\left(\pm \frac{{\rm i}\Gamma_o}{2}\psi_{2i-1}\psi_{2i}\right)$.
According to Born's rule, the probability $p_{i;\pm}$ of obtaining outcome $\pm$ is
\begin{equation}
\label{eq: app p cosh2}
p_{i;\pm}
=
\frac{1}{2}\pm \frac{1}{2}\langle {\rm i}\psi_{2i-1}\psi_{2i}\rangle\,\tanh \Gamma_o\, .
\end{equation}
The single-particle representation of $\mK_{2i-1,\pm}$ is $K_{2i-1,\pm}=e^{\pm H}$, where $H$
is given as $H|_{\{2i-1,2i\}}={\rm i}\Gamma_o\begin{pmatrix}0&1\\-1&0\end{pmatrix}$ and vanishes on all the other sites.
After applying the Kraus operator, the updated state $|\Psi_1\rangle$ is annihilated by
$c_n'=\frac{1}{\sqrt{2}}\sum_{i=1}^{L}\psi_i\,(K_{2i-1,\pm}U)_{in}$.
We then perform a QR decomposition of the $L\times (L/2)$ matrix $K_{2i-1,\pm}U$:
$K_{2i-1,\pm}U=U_1 R_1$,
where $U_1^{\dagger}U_1=1_{L/2\times L/2}$ and $R_1$ is upper triangular.
Since the columns of $U_1$ span the same subspace as those of $K_{2i-1,\pm}U$, the state $|\Psi_1\rangle$ is equivalently annihilated by
$c_n''=\frac{1}{\sqrt{2}}\sum_{i=1}^{L}\psi_i\,(U_1)_{in}$.
The equal-time correlation matrix of the updated Gaussian state can then be computed from $U_1$.
$\langle \Psi_1|\,{\rm i}[\psi_i,\psi_j]\,|\Psi_1\rangle$ is given by $-2\ii U_1U_1^{\dagger} + \ii$, which in turn fixes the Born probabilities for subsequent measurements.

We repeat this procedure after each application of a single-particle Kraus operator $K_1,K_2,\ldots,K_N$.
After $N$ steps, the evolution can be written as
$K_N\cdots K_1\,U = U_N R_N R_{N-1}\cdots R_1$.
For $N\gg 1$, the Lyapunov exponents are extracted from the accumulated triangular factors as
$\eta_i = \frac{1}{N}\sum_{j=1}^{N}\ln |(R_j)_{ii}|$ for $i=1,2,\ldots,L/2$.
Due to particle-hole symmetry, the remaining $L/2$ exponents satisfy $\eta_{L+1-i}=-\eta_i$.

The algorithm is similar even in the presence of unitary dynamics. 
We multiply the state matrix by the single-particle unitary evolution operator.
Since the resulting state vector is still unitary, we do not need an additional QR stabilization step.
For complex fermions, the procedure is analogous, with the state represented by an orthonormal set of occupied single-particle orbitals and updated under the corresponding single-particle Kraus and unitary operators.
Further implementation details can be found in, e.g., Refs.~\cite{cao2019,zhenyuxiao2024}.

\bibliography{ref}

\end{document}